\documentclass[11pt]{article}
\usepackage{epsfig,amsmath,epic,eepic}
\input amssym.def
\input amssym.tex

\newcommand{\qed}{\rule{3mm}{3mm}}
\newcommand{\itbf}{\itshape\bfseries}

\newcommand{\ee}{{\frak e}}

\newcommand{\cD}{{\cal D}}
\newcommand{\cG}{{\cal G}}

\newcommand{\cL}{{\cal L}}

\newcommand{\wx}{\widetilde{x}}

\newbox\meibox
\def\placeunder#1#2#3#4{\setbox\meibox%
\vbox{\hbox{\hskip#4$\hphantom{#2}$}\hbox{$\hphantom{#1}$}}%
\vtop{\baselineskip=0pt\lineskiplimit=\baselineskip%
\lineskip=#3\hbox to \wd\meibox{\hfil\hskip#4$#2$\hfil}%
\hbox to \wd\meibox{\hfil$#1$\hfil}}}
\def\undertilde#1{\mathchoice{%
\placeunder{\vbox to 1.4pt{\hbox{$\displaystyle\widetilde{\,\,\,
}$}\vss}}{\displaystyle#1}{1.5pt}{1.5pt}}%
{\placeunder{\vbox to 1.4pt{\hbox{$\textstyle\widetilde{\,\,
}$}\vss}}{\textstyle#1}{1.5pt}{1.5pt}}%
{\placeunder{\vbox to 1.4pt{\hbox{$\scriptstyle\tilde{
}$}\vss}}{\scriptstyle#1}{1pt}{1pt}}%
{\placeunder{\vbox to 1.4pt{\hbox{$\scriptscriptstyle\tilde{
}$}\vss}}{\scriptscriptstyle#1}{1pt}{1pt}}%
}

\newtheorem{theorem}{Theorem}
\newtheorem{proposition}[theorem]{Proposition}
\newtheorem{lemma}[theorem]{Lemma}
\newtheorem{corollary}[theorem]{Corollary}
\newtheorem{definition}[theorem]{Definition}

\newtheorem{remark}[theorem]{Remark}

\begin{document}

\begin{center}{\LARGE\bf Integrable systems on quad-graphs}\end{center}
\begin{center}{\Large A.I.\,Bobenko\footnote{E--mail: {\tt
bobenko@math.tu-berlin.de}}, Yu.B.\,Suris\footnote{E--mail: {\tt
suris@sfb288.math.tu-berlin.de}}}\end{center}
\begin{center}{\it Institut f\"ur Mathematik, Fachbereich II, TU Berlin, \\
Str. des 17. Juni 136, 10623 Berlin, GERMANY}\end{center}
\vspace{5mm}

{\small {\bf Abstract.} We consider general integrable systems on graphs
as discrete flat connections with the values in loop groups. We argue that
a certain class of graphs is of a special importance in this respect, namely
quad-graphs, the cellular decompositions of oriented surfaces with all
two--cells being quadrilateral. We establish a relation between integrable
systems on quad-graphs and discrete systems of the Toda type on graphs. We
propose a simple and general procedure for deriving discrete zero 
curvature representations for integrable systems on quad-graphs, based on
the principle of the three--dimensional consistency. Thus, finding a zero
curvature representation is put on an algorithmic basis and does not rely
on the guesswork anymore. Several examples of integrable systems on quad-graphs
are considered in detail, their geometric interpretation is given in terms of
circle patterns.}

\setcounter{equation}{0}
\section{Introduction}

Discrete (lattice) systems constitute a well--established part of the theory 
of integrable systems. They came up already in the early days of the theory,
see, e.g. \cite{H1}, \cite{H2}, and took gradually more and more important
place in it, cf. a review in \cite{NC}. Nowadays many experts in the field
agree that discrete integrable systems are in many respects even more 
fundamental than the continuous ones. They play a prominent role in various
applications of integrable systems such as discrete differential geometry,
see, e.g., a review in \cite{BP3}.

Traditionally, independent variables of discrete integrable systems are
considered as belonging to a regular square lattice ${\Bbb Z}^2$ (or
its multi-dimensional analogs ${\Bbb Z}^d$). Only very recently, there
appeared first hints on the existence of a rich and meaningful theory
of integrable systems on non--square lattices and, more generally, on
arbitrary graphs. The relevant publications are
almost exhausted by \cite{ND}, \cite{OP}, \cite{A2}, \cite{NS}, \cite{KN2},
\cite{BHS}, \cite{BH}, \cite{A3}. 

We define integrable systems on graphs as flat connections with the values
in loop groups. This is is very natural definition, and experts in
discrete integrable systems will not only immediately accept it, but might
even consider it trivial. Nevertheless, it crystallized only very recently,
and seems not to appear in the literature before \cite{BHS}, \cite{BH},
\cite{A3}. (It should be noted that a different framework for integrable 
systems on graphs is being developed by S.P.Novikov with collaborators 
\cite{ND}, \cite{NS}, \cite{KN2}.) 
  
We were led to considering such systems by our (with T.\,Hoffmann) 
investigations of circle patterns as objects of discrete complex
analysis: in Refs. \cite{BHS}, \cite{BH} we demonstrated that certain classes
of circle patterns with the combinatorics of regular hexagonal lattice
are directly related to integrable systems. To realize this, it was necessary
to formulate the notion of an integrable system on a graph which is not a
square lattice (the regular hexagonal lattice is just one example of such 
a graph). 

Another context where such systems appear naturally is the
differential geometry of discretized surfaces, see \cite{BP3}. It was
realized in the latter reference that natural domains where discrete
analogs of parametrized surfaces are defined are constituted by {\it
quad-graphs}, i.e. cellular decompositions of surfaces whose two--cells
are all quadrilaterals (see Fig.1 in \cite{BP3}). However, a systematical
study of integrable systems on quad-graphs was not put forward in \cite{BP3}.

Finally, one of the sources of the present development is the theory
of integrable discretizations of one--dimensional lattice systems of the
relativistic Toda type \cite{S}. It was demonstrated by V.\,Adler \cite{A2} 
that these discretizations are naturally regarded as integrable systems on 
the regular triangular lattice, which eventually led him to the discovery of 
systems of the Toda type on arbitrary graphs \cite{A3}.

In the present paper we put forward the the viewpoint according to which 
integrable systems on quad-graphs are very fundamental and explain many
known phenomena in the field. In particular, we shall present the {\it
derivation} of zero curvature representations with a spectral parameter
for such systems from a very simple and fundamental principle, namely
the consistency on a three--dimensional quad-graph. It should be noted
that usually the zero curvature (or Lax) representation is considered as
a rather transcendental attribute of a given integrable system, whose
existence is easy to verify but very difficult to guess just by looking at the
equation. Our approach shows how to derive a zero curvature representation
for a given discrete system provided it possess the property of the
three--dimensional consistency. Taking into account that, as widely believed, 
the whole universe of two--dimensional continuous integrable systems
(described by differential--difference or partial differential equations)
may be obtained from the discrete ones via suitable limit procedures,
we come to the conclusion that there is nothing mysterious in zero
curvature representations anymore: it can be found in a nearly algorithmic way. 
Also, we solve a lesser mystery of the existence of Toda type systems on 
arbitrary graphs \cite{A3}, in that we trace there origin back to much more 
simple and fundamental systems on quad-graphs.

The structure of the paper is as follows: in Sect. \ref{Sect integrable systems 
on graphs} we present the definition of integrable systems on graphs in
more details. In Sect. \ref{Sect quad-graphs} we consider quad-graphs and
discuss their relation to more general graphs. Sect. \ref{Sect from diamond to 
Toda} contains generalities on integrable systems on quad-graphs (four--point
equations) and their relation to the discrete Toda type systems (equations
on stars). Sect. \ref{Sect cross-ratio A}--\ref{Sect cross-ratio D}
are devoted to a detailed account of several concrete integrable systems
on quad-graphs, namely the {\it cross-ratio system} and two of its
generalizations. This account includes the algorithmic derivation of the
zero curvature representation, discussion of the duality and related
constructions, derivation of the corresponding systems of the Toda type
along with their zero curvature representations, and a geometric interpretation
in terms of circle patterns. Finally, conclusions and perspectives for the 
further research are given in Sect. \ref{Sect conclusions}.

\setcounter{equation}{0}
\section{Discrete flat connections on graphs}
\label{Sect integrable systems on graphs}

We consider ``integrable systems on graphs'' as flat connections with the
values in loop groups. More precisely, this notion includes the following
ingredients:
\begin{itemize}
\item  A {\itbf cellular decomposition} $\cG$ of an oriented surface. 
The set of its vertices will be denoted $V(\cG)$, the set of its 
edges will be denoted $E(\cG)$, and the set of its faces will be denoted 
$F(\cG)$. For each edge, one of its possible orientations is fixed.
\item A {\itbf loop group} $G[\lambda]$, whose elements are functions from
${\Bbb C}$ into some group $G$. The complex argument $\lambda$ of
these functions is known in the theory of integrable systems as
the ``spectral parameter''.
\item A {\itbf ``wave function''} $\Psi: V(\cG)\mapsto G[\lambda]$,
defined on the vertices of $\cG$.
\item A collection of {\itbf ``transition matrices''} 
$L: E(\cG)\mapsto G[\lambda]$ defined on the edges of $\cG$.
\end{itemize}
It is supposed that for any oriented edge $\ee=(v_1,v_2)\in E(\cG)$ 
the values of the wave functions in its ends are connected via
\begin{equation}\label{wave function evol}
\Psi(v_2,\lambda)=L(\ee,\lambda)\Psi(v_1,\lambda).
\end{equation}
Therefore the following {\itbf discrete zero curvature condition} is
supposed to be satisfied. Consider any closed contour consisting
of a finite number of edges of $\cG$:
\[
\ee_1=(v_1,v_2),\quad \ee_2=(v_2,v_3),\quad \ldots,\quad \ee_n=(v_n,v_1).
\]
Then
\begin{equation}\label{zero curv cond}
L(\ee_n,\lambda)\cdots L(\ee_2,\lambda)L(\ee_1,\lambda)=I.
\end{equation}
In particular, for any edge $\ee=(v_1,v_2)$, if $\ee^{-1}=(v_2,v_1)$, then
\begin{equation}\label{zero curv cond inv}
L(\ee^{-1},\lambda)=\Big(L(\ee,\lambda)\Big)^{-1}.
\end{equation}

Actually, in applications the matrices $L(\ee,\lambda)$ depend
also on a point of some set $X$ (the ``phase space'' of an
integrable system), so that some elements $x(\ee)\in X$ are
attached to the edges $\ee$ of $\cG$. In this case the discrete
zero curvature condition (\ref{zero curv cond}) becomes equivalent
to the collection of equations relating the fields $x(\ee_1)$,
$\ldots$, $x(\ee_p)$ attached to the edges of each closed contour.
We say that this collection of equations admits a {\itbf zero
curvature representation}.

For an arbitrary graph, the analytical consequences of the zero
curvature representation for a given collection of equations are
not clear. However, in case of regular graphs, like those generated
by the square lattice ${\Bbb Z}+i{\Bbb Z}\subset{\Bbb C}$, or by the regular 
triangular lattice ${\Bbb Z}+e^{2\pi i/3}{\Bbb Z}\subset{\Bbb C}$,
such representation may be used to determine conserved quantities
for suitably defined Cauchy problems, as well as to apply powerful
analytical methods for finding concrete solutions.

\setcounter{equation}{0}
\section{Quad-graphs}
\label{Sect quad-graphs}

Our point in this paper is that, although one can consider 
integrable systems on very different kinds of graphs, there is one kind
-- quad-graphs -- supporting the most fundamental integrable systems.
\begin{definition}
A cellular decomposition $\cG$ of an oriented surface is called a
{\itbf quad-graph}, if all its faces are quadrilateral.
\end{definition}

Before we turn to illustrating this claim by concrete examples, we would like 
to give a construction which produces from an arbitrary cellular
decomposition a certain quad-graph. Towards this aim, we first recall 
the notion of the {\it dual graph}, or, more precisely, of the {\itbf dual 
cellular decomposition} $\cG^*$. The vertices from $V(\cG^*)$ are in one-to-one 
correspondence to the faces from $F(\cG)$ (actually, they can be chosen as some 
points inside the corresponding faces, cf. Fig. \ref{dual vertex}). 
Each $\ee\in E(\cG)$ separates two faces from $F(\cG)$, which in turn
correspond to two vertices from $V(\cG^*)$. A path between these two vertices
is then declared as an edge $\ee^*\in E(\cG^*)$ dual to $\ee$. 
Finally, the faces from $F(\cG^*)$ are in a one-to-one correspondence with
the vertices from $V(\cG)$: if $v_0\in V(\cG)$, and $v_1,\ldots,v_n\in V(\cG)$ 
are its neighbors connected with $v_0$ by the edges $\ee_1=(v_0,v_1),\ldots
\ee_n=(v_0,v_n)\in E(\cG)$, then the face from $F(\cG^*)$ corresponding to
$v_0$ is defined by its boundary $\ee_1^*\cup\ldots\cup\ee_n^*$ (cf. Fig. 
\ref{dual face}).

\begin{figure}[htbp]
    \setlength{\unitlength}{28.45pt}
    \begin{minipage}[t]{200pt}
\begin{picture}(6,5)
\put(3,1){\circle*{0.2}}
\put(4.5,1.5){\circle*{0.2}}
\put(5,2.7){\circle*{0.2}}
\put(3.5,4){\circle*{0.2}}
\put(2.1,2.5){\circle*{0.2}}
\put(3.7,2.6){\circle{0.2}}
\path(3,1)(4.5,1.5)  \path(3,1)(2.8,0.6)
\path(3,1)(3.2,0.6)  \path(4.5,1.5)(5,2.7)
\path(4.5,1.5)(4.9,1.3)
\path(5,2.7)(3.5,4)  \path(5,2.7)(5.3,2.4)  \path(5,2.7)(5.3,3)
\path(3.5,4)(2.1,2.5)  \path(3.5,4)(3.5,4.4)
\path(2.1,2.5)(3,1)  \path(2.1,2.5)(1.7,2.5)
\path(2.1,2.5)(1.8,2.3)  \path(2.1,2.5)(1.8,2.7)
\dashline[+30]{0.2}(3.7,2.6)(4.9,4.2)
\dashline[+30]{0.2}(3.7,2.6)(5.5,1.7)
\dashline[+30]{0.2}(3.7,2.6)(3.9,0.6)
\dashline[+30]{0.2}(3.7,2.6)(2.2,1.5)
\dashline[+30]{0.2}(3.7,2.6)(2.4,3.6)
\end{picture}
    \caption{The vertex from $V(\cG^*)$ dual to the face from $F(\cG)$.
    }
        \label{dual vertex}
	\end{minipage}\hfill
\begin{minipage}[t]{200pt}
\begin{picture}(6,5)
\put(3.5,2.6){\circle*{0.2}}
\put(6,2.4){\circle*{0.2}}
\put(4.7,4.5){\circle*{0.2}}
\put(2.3,4.2){\circle*{0.2}}
\put(1,2.4){\circle*{0.2}}
\put(2.4,0.8){\circle*{0.2}}
\put(4.8,0.6){\circle*{0.2}}
\path(3.5,2.6)(6,2.4)   \path(3.5,2.6)(4.7,4.5)
\path(3.5,2.6)(2.3,4.2)  \path(3.5,2.6)(1,2.4)
\path(3.5,2.6)(2.4,0.8)  \path(3.5,2.6)(4.8,0.6)
\path(6,2.4)(6.3,2.6)  \path(6,2.4)(6.3,2.2)
\path(4.7,4.5)(4.7,4.9) \path(4.7,4.5)(5.1,4.7)
\path(2.3,4.2)(2.2,4.6)   \path(2.3,4.2)(1.9,4.3)
\path(1,2.4)(1,2.8)  \path(1,2.4)(0.9,2.1)
\path(2.4,0.8)(2,0.8)  \path(2.4,0.8)(2.4,0.4)
\path(4.8,0.6)(5.2,0.5)  \path(4.8,0.6)(4.8,0.2)
\put(4.9,3.3){\circle{0.2}}
\put(3.5,3.9){\circle{0.2}}
\put(2.1,3.2){\circle{0.2}}
\put(2.3,1.8){\circle{0.2}}
\put(3.6,1.2){\circle{0.2}}
\put(4.9,1.7){\circle{0.2}}
\dashline[+30]{0.2}(4.9,3.3)(3.5,3.9)
\dashline[+30]{0.2}(3.5,3.9)(2.1,3.2)
\dashline[+30]{0.2}(2.1,3.2)(2.3,1.8)
\dashline[+30]{0.2}(2.3,1.8)(3.6,1.2)
\dashline[+30]{0.2}(3.6,1.2)(4.9,1.7)
\dashline[+30]{0.2}(4.9,1.7)(4.9,3.3)
\end{picture}
    \caption{The face from $F(\cG^*)$ dual to the vertex from $V(\cG)$.}
        \label{dual face}
	\end{minipage}
\end{figure}
\vspace{1cm} 

\begin{figure}[htbp]
\begin{center}
\setlength{\unitlength}{0.05em}
\begin{picture}(200,240)(-100,-110)
 \put(-100,0){\circle*{10}}\put(100,0){\circle*{10}}
 \put(0,-100){\circle{10}} \put(0,100){\circle{10}}
 \path(-100,0)(100,0)
 \path(-100,0)(-120,70)
 \path(-100,0)(-120,-70)
 \path(100,0)(120,70)
 \path(100,0)(120,-70)
 \dashline[+30]{10}(0,100)(0,-100)
 \dashline[+30]{10}(0,100)(-140, 30)
 \dashline[+30]{10}(0,100)(140, 30)
 \dashline[+30]{10}(0,-100)(-140, -30)
 \dashline[+30]{10}(0,-100)(140, -30)
 \thicklines
 \path(-100,0)(0,-100)
 \path(0,-100)(100,0)
 \path(100,0)(0,100)
 \path(0,100)(-100,0)
 \put(-125,  -7){$v_1$} \put(110, -7){$v_2$}
 \put(  -6,-125){$f_1$} \put( -6,115){$f_2$}
\end{picture}
\caption{A face of the double}\label{diamond}
\end{center}
\end{figure}
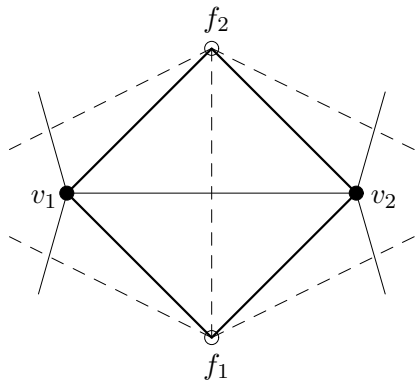

Now, following \cite{M}, we introduce a new complex, the {\itbf double} $\cD$,
constructed from $\cG$, $\cG^*$ (notice that our terminology here is different
from that of \cite{M}: the complex $\cD$ is called a {\it ``diamond''} there,
while the term ``double'' is used for another object). The set of vertices of
the double $\cD$ is $V(\cD)=V(\cG)\cup V(\cG^*)$. 
Each pair of dual edges, say $\ee=(v_1,v_2)$ and $\ee^*=(f_1,f_2)$, 
as on Fig.\,\ref{diamond}, defines a quadrilateral
$(v_1,f_1,v_2,f_2)$, and all these quadrilaterals constitute the faces of the
cell decomposition (quad-graph) $\cD$. Let us stress that the edges of $\cD$ 
belong neither to $E(\cG)$ nor to $E(\cG^*)$. See Fig.\,\ref{diamond}.

Quad-graphs $\cD$ coming as doubles have the following property:
the set $V(\cD)$ may be decomposed into two complementary halves, 
$V(\cD)=V(\cG)\cup V(\cG^*)$ (``black'' and ``white'' vertices), such that 
two ends of each edge from $E(\cD)$ are of different colours. Equivalently,
any closed loop consisting of edges of $\cD$ has an even length. We shall call
such quad-grahs {\itbf even}. 

Conversely, any even quad-graph $\cD$ may be 
considered as a double of some cellular decomposition $\cG$. The edges from 
$E(\cG)$, say, are defined then as paths joining two ``black'' vertices of 
each face from $F(\cD)$. (This decomposition of $V(\cD)$ into $V(\cG)$ and 
$V(\cG^*)$ is unique, up to interchanging the roles of $\cG$ and $\cG^*$.)
 
Notice that if $\cD$ is not even (i.e. if it admits loops consisting of an 
odd number of edges), then one can easily produce from $\cD$ a new even 
quad-graph $\cD'$, simply by refining each of the quadrilaterals from $F(\cD)$ 
into four smaller ones.  

In the present paper we are intereted mainly in the local theory of integrable
systems of quad--graphs. Therefore, in order to avoid global considerations, 
we always assume (without mentioning it explicitly) that our quad-graphs are 
cellular decompositions of a topological disc (open or closed). In particular,
our quad-graphs $\cD$ are always even, so that $\cG$ and $\cG^*$ are 
well-defined.
\vspace{3mm}

We shall consider in this paper several integrable systems on quad-graphs.
For all these systems the ``fields'' $z:V(\cD)\mapsto\widehat{\Bbb C}$ will be 
attached to the {\it vertices} rather than to the edges of the graph, so that 
for an edge $\ee=(v_0,v_1)\in E(\cD)$ the transition matrix will be written as 
\[
L(\ee,\lambda)=L(z_1,z_0,\lambda).
\]
Here and below $z_k=z(v_k)$. It is easy to understand that the condition 
(\ref{zero curv cond inv}) for the transition matrices reads in the present 
set-up as
\begin{equation}\label{zero curv inv prelim}
L(z_0,z_1,\lambda)=\Big(L(z_1,z_0,\lambda)\Big)^{-1},
\end{equation}
while the discrete zero curvature condition (\ref{zero curv cond}) in the 
case of quad-graphs is equivalent to 
\begin{equation}\label{quad-gr Lax prelim}
L(z_2,z_1,\lambda)L(z_1,z_0,\lambda)=L(z_2,z_3,\lambda)L(z_3,z_0,\lambda).
\end{equation}
In all our examples this matrix equation will be equivalent to a single
scalar equation
\begin{equation}\label{eq preprelim}
\Phi(z_0,z_1,z_2,z_3)=0,
\end{equation}
relating four fields sitting on the four vertices of an arbitrary face
from $F(\cD)$. Moreover, in all our examples it will be possible to uniquely
solve the equation (\ref{eq prelim}) for any field $z_0,\ldots,z_3$ in terms
of other three ones.

\setcounter{equation}{0}
\section{Equations on quad-graphs and discrete systems of the Toda type}
\label{Sect from diamond to Toda} 

Of course, the above constructions might seem well-known, and there are 
several four-point integrable equations of the type (\ref{eq prelim}) 
available in the literature, see, e.g., \cite{H1}, \cite{H2}, \cite{NQC}, 
\cite{NC}, \cite{FV}, \cite{BP1}, \cite{BP2}, \cite{A1}. However, these 
equations were always considered only on the 
simplest possible quad-graph, namely on the square lattice $V(\cD)={\Bbb Z}^2$. 
We will see, however, that the novel feature introduced in the present paper, 
namely considering quad-graphs of arbitrary combinatorics, has interesting and
far reaching consequences. One of such consequences is the derivation of 
{\it equations of the discrete Toda type on arbitrary graphs} which were first
introduced in \cite{A3}. 

\begin{figure}[htbp]
\begin{center}
\setlength{\unitlength}{0.05em}
\begin{picture}(200,240)(-100,-110)
 \put(-100,0){\circle*{10}}\put(100,0){\circle*{10}}
 \put(0,-100){\circle{10}} \put(0,100){\circle{10}}
 \thicklines
 \path(-100,0)(0,-100)
 \path(0,-100)(100,0)
 \path(100,0)(0,100)
 \path(0,100)(-100,0)
 \put(-125,  -7){$v_0$} \put(110, -7){$v_2$}
 \put(  -6,-125){$v_1$} \put( -6,115){$v_3$}
 \put(-64,60){$\alpha_2$} \put(60,-60){$\alpha_2$}
 \put(57,57){$\alpha_1$} \put(-65,-64){$\alpha_1$}
\end{picture}
\caption{A face of the labelled quad-graph}\label{diamond again}
\end{center}
\end{figure}
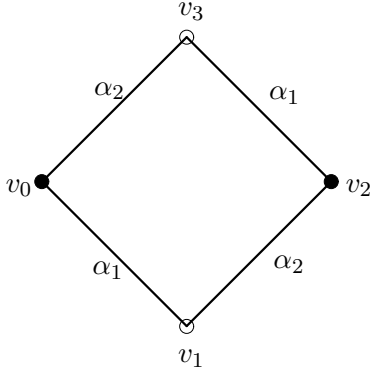

At this point we have to specify the set-up further. We shall assume that
the transition matrices depend additionally on a {\it parameter} 
$\alpha=\alpha(\ee)$ assigned to each (non-oriented) edge:
\[
L(\ee,\lambda)=L(z_1,z_0,\alpha,\lambda).
\]
So, actually, an integrable system will be characterized by a function 
$\alpha: E(\cD)\mapsto{\Bbb C}$. For some reasons which will become clear soon, 
the function $\alpha$ will always satisfy an additional condition.
\begin{definition}\label{labelling}
A {\itbf labelling} of $E(\cD)$ is a function $\alpha: E(\cD)\mapsto {\Bbb C}$
satisfying the following condition: the values of $\alpha$ on two opposite
edges of any quadrilateral from $F(\cD)$ are equal to one another. 
\end{definition}
This is illustrated on Fig.\,\ref{diamond again}. Obviously, there exist 
infinitely many labellings, all of them may be 
constructed as follows: choose some value of $\alpha$ for an arbitrary edge 
of $\cD$, and assign consecutively the same value to all ``parallel'' edges 
along a strip of quadrilaterals, according to the definition of labelling. 
After that, take an arbitrary edge still without a label, choose some value
of $\alpha$ for it, and extend the same value along the corresponding
strip of quadrilaterals. Proceed similarly, till all edges of $\cD$ are
exhausted. 

The condition (\ref{zero curv inv prelim}) for
the transition matrices reads in the present set-up as
\begin{equation}\label{zero curv inv}
L(z_0,z_1,\alpha,\lambda)=\Big(L(z_1,z_0,\alpha,\lambda)\Big)^{-1},
\end{equation}
while the discrete zero curvature condition (\ref{quad-gr Lax prelim}) 
in the notations of Fig.\,\ref{diamond again} reads as 
\begin{equation}\label{quad-gr Lax}
L(z_2,z_1,\alpha_2,\lambda)L(z_1,z_0,\alpha_1,\lambda)=
L(z_2,z_3,\alpha_1,\lambda)L(z_3,z_0,\alpha_2,\lambda).
\end{equation}
The left--hand side of the scalar equation (\ref{eq preprelim}) also
depends on the parameters:
\begin{equation}\label{eq prelim}
\Phi(z_0,z_1,z_2,z_3,\alpha_1,\alpha_2)=0.
\end{equation}
We shall require two additional features of the equation (\ref{eq prelim}). 
The first one is quite natural and is satisfied by all known examples: 
the function $\Phi$ is symmetric with respect to the cyclic shift 
$(z_0,z_1,z_2,z_3)\mapsto(z_1,z_2,z_3,z_0)$ accompanied by
interchanging the parameters $\alpha_1\leftrightarrow\alpha_2$. The second
feature might seem not natural at all, at least on the first sight, but also 
holds, for some mysterious reasons, for the majority of (if not all) the known 
examples. Namely, suppose that the equation (\ref{eq prelim}) may be 
equivalently rewritten in the form
\begin{equation}\label{eq 3-leg}
f(z_0,z_1,\alpha_1)-f(z_0,z_3,\alpha_2)=g(z_0,z_2,\alpha_1,\alpha_2).
\end{equation}
We call the equation (\ref{eq 3-leg}) the 
{\itbf three--leg form} of the equation (\ref{eq prelim}). Indeed, its three 
terms may be associated with three ``legs'' $(v_0,v_1)$, $(v_0,v_3)$, and 
$(v_0,v_2)$. Notice that while the first two ``legs'' belong to $E(\cD)$, 
the third one does not; instead, it belongs to $E(\cG^*)$ (recall that our
quad-graph $\cD$ comes as a double constructed from $\cG$, $\cG^*$). 
Now consider $n$ faces from $F(\cD)$ having common ``black'' vertex $v_0\in
V(\cG)\subset V(\cD)$ (cf. Fig.\,\ref{flower}, where $n=5$).
\begin{figure}[htbp]
    \setlength{\unitlength}{20pt}
    \begin{minipage}[t]{180pt}
\begin{picture}(8,9)
\put(4,4){\circle*{0.2}}
\put(4.2,8){\circle*{0.2}}
\put(5,6){\circle{0.2}}
\put(3,5.7){\circle{0.2}}
\put(7,5.9){\circle*{0.2}}
\put(6,4.2){\circle{0.2}}
\put(7.5,2.6){\circle*{0.2}}
\put(5.1,2.5){\circle{0.2}}
\put(4,0.2){\circle*{0.2}}
\put(3,2.1){\circle{0.2}}
\put(0.4,3.9){\circle*{0.2}}
\thicklines
\path(4,4)(5,6) \path(5,6)(4.2,8) \path(4.2,8)(3,5.7) \path(3,5.7)(4,4)
\path(3,5.7)(0.4,3.9) \path(0.4,3.9)(3,2.1) \path(3,2.1)(4,4)
\path(3,2.1)(4,0.2) \path(4,0.2)(5.1,2.5) \path(5.1,2.5)(4,4)
\path(5.1,2.5)(7.5,2.6) \path(7.5,2.6)(6,4.2) \path(6,4.2)(4,4)
\path(6,4.2)(7,5.9) \path(7,5.9)(5,6)
\put(3.3,4){$v_0$}
\put(6.3,4.2){$v_1$}  \put(7.3,5.9){$v_2$}  \put(5.1,6.3){$v_3$}
\put(4.1,8.3){$v_4$}  \put(2.5,6){$v_5$}    \put(0.2,4.2){$v_6$}
\put(2.6,1.7){$v_7$}  \put(3.9,-0.3){$v_8$} \put(5.1,2.1){$v_9$}
\put(7.5,2.2){$v_{10}$}
\put(4.8,4.3){$\alpha_1$}  \put(4.7,5){$\alpha_2$}
\put(6.7,5){$\alpha_2$} \put(6.1,6.2){$\alpha_1$}
\put(3.5,5.1){$\alpha_3$}  \put(3.5,2.7){$\alpha_4$}
\put(4.7,3.2){$\alpha_5$}
\end{picture}
    \caption{Faces from $F(\cD)$ around the vertex $v_0$.}
        \label{flower}
	\end{minipage}\hfill
    \begin{minipage}[t]{180pt}
\begin{picture}(8,9)
\put(4,4){\circle*{0.2}}
\put(4.2,8){\circle*{0.2}}
\put(7,5.9){\circle*{0.2}}
\put(7.5,2.6){\circle*{0.2}}
\put(4,0.2){\circle*{0.2}}
\put(0.4,3.9){\circle*{0.2}}
\path(4,4)(4.2,8) \path(4,4)(7,5.9) \path(4,4)(7.5,2.6) 
\path(4,4)(4,0.2) \path(4,4)(0.4,3.9) 
\put(3.1,4.4){$v_0$} 
\put(7.3,5.9){$v_2$}  
\put(4.1,8.3){$v_4$}  \put(0.2,4.2){$v_6$}
\put(3.9,-0.3){$v_8$} \put(7.5,2.2){$v_{10}$}
\put(4.5,4){$\alpha_1$}  \put(4.1,4.5){$\alpha_2$}
\put(6.7,5.3){$\alpha_2$} \put(6.1,5.9){$\alpha_1$}
\put(3.3,3.5){$\alpha_4$}
\put(4.1,3.3){$\alpha_5$}
\end{picture}
    \caption{The star (in $\cG$) of the vertex $v_0$.}
        \label{star}
	\end{minipage}
	\end{figure}
Summing up the equations (\ref{eq 3-leg}) for these $n$ faces, i.e.
\begin{equation}\label{eq 3-leg gen}
f(z_0,z_{2k-1},\alpha_k)-f(z_0,z_{2k+1},\alpha_{k+1})=
g(z_0,z_{2k},\alpha_k,\alpha_{k+1}),
\end{equation}
we end up with the equation which includes only the fields $z$ in the 
``black'' vertices:
\begin{equation}\label{Toda type}
\sum_{k=1}^n g(z_0,z_{2k},\alpha_k,\alpha_{k+1})=0.
\end{equation}
The latter formula may be interpreted as a system on the graph $\cG$:
there is one equation per vertex $v_0\in V(\cG)$, and this equation relates 
the fields on the {\it star} of $v_0$ consisting of the edges from $E(\cG)$ 
incident to $v_0$, see Fig.\,\ref{star}. The parameters $\alpha_k$ are then
viewed as assigned to the corners of the faces of $\cG$. Observe that 
Definition \ref{labelling} of a labelling of $E(\cD)$ may be reformulated  
in terms of $\cG$ alone as follows: the corner parameters of two faces 
adjacent to the edge $(v_0,v_{2k})\in E(\cG)$ at the vertex $v_0$ 
are the same as at the vertex $v_{2k}$ (the ordering of the parameters at
$v_0$ when looking towards $v_{2k}$ is the same as the ordering of the 
parameters at $v_{2k}$ when looking towards $v_0$, cf. Fig.\,\ref{star}). 

The collection of equations (\ref{Toda type}) constitutes the 
{\it discrete Toda type system on $\cG$}. Obviously, the nature of this 
system is formally somewhat different from the general construction of
Sect. \ref{Sect integrable systems on graphs} (the former relates fields
on stars, while the latter relates fields on closed contours). However,
our construction suggests the following type of a zero curvature
representation for (\ref{Toda type}). The wave function $\Psi:V(\cG^*)\mapsto 
G[\lambda]$ is defined on the ``white'' vertices $v_{2k-1}$ (in the notations
of Fig. \ref{flower}), i.e. on the vertices of the dual graph. The transition 
matrix along the edge $\ee^*=(v_{2k-1},v_{2k+1})\in E(\cG^*)$ of the dual 
graph is equal to
\[
L(z_{2k+1},z_{2k},\alpha_k,\lambda)L(z_{2k}.z_{2k-1},\alpha_{k+1},\lambda)=
L(z_{2k+1},z_0,\alpha_{k+1},\lambda)L(z_0,z_{2k-1},\alpha_k,\lambda).
\]
If one prefers to consider everything in terms of $\cG$ alone, one can think
of $\Psi$ as defined on the faces of $\cG$, and of the above matrix 
as corresponding to the transition {\it across} the edge
$\ee=(v_0,v_{2k})\in E(\cG)$. One would like this matrix to depend only
on the fields $z_0$, $z_{2k}$. In all our examples this can and will be 
achieved by a certain gauge transformation $\Psi(v_k)\mapsto A(v_k)\Psi(v_k)$.

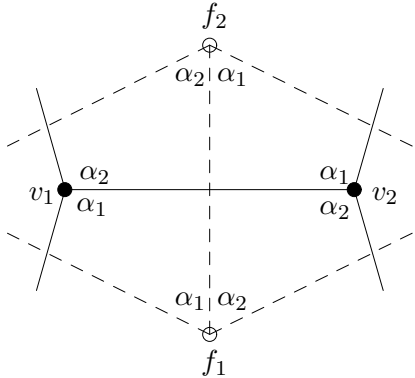
\begin{figure}[htbp]
\begin{center}
\setlength{\unitlength}{0.05em}
\begin{picture}(200,240)(-100,-110)
 \put(-100,0){\circle*{10}}\put(100,0){\circle*{10}}
 \put(0,-100){\circle{10}} \put(0,100){\circle{10}}
 \path(-100,0)(100,0)
 \path(-100,0)(-120,70)
 \path(-100,0)(-120,-70)
 \path(100,0)(120,70)
 \path(100,0)(120,-70)
 \dashline[+30]{10}(0,100)(0,-100)
 \dashline[+30]{10}(0,100)(-140, 30)
 \dashline[+30]{10}(0,100)(140, 30)
 \dashline[+30]{10}(0,-100)(-140, -30)
 \dashline[+30]{10}(0,-100)(140, -30)
 \put(-90,  7){$\alpha_2$} \put(76,  7){$\alpha_1$}
 \put(-92,-15){$\alpha_1$}\put(76,-17){$\alpha_2$}
 \put(-24, 75){$\alpha_2$} \put( 5, 75){$\alpha_1$}
 \put(-24,-80){$\alpha_1$}\put( 5,-80){$\alpha_2$}
 \put(-125,  -7){$v_1$} \put(112, -7){$v_2$}
 \put(  -6,-125){$f_1$} \put( -6,115){$f_2$}
\end{picture}
\caption{Corner parameters of the dual graph}\label{fig.dual}
\end{center}
\end{figure}

Next, notice that in the above reasoning the roles of $\cG$ and $\cG^*$ may 
be interchanged. Hence, one can derive a Toda type system on $\cG^*$ as well.
The distribution of the corner parameters on $\cG$ and $\cG^*$ is illustrated
on Fig.\,\ref{fig.dual}.

We arrive at the following statement.
\begin{proposition}\label{quadgraph to Toda}
a) Let the system (\ref{eq prelim}) on a quad-graph $\cD$ admit a three-leg 
form. Any solution $z:V(\cD)\mapsto{\Bbb C}$ of this system, 
restricted to $V(\cG)$, resp. to $V(\cG^*)$, delivers a solution of the 
discrete Toda type system (\ref{Toda type}) on $\cG$, resp. on $\cG^*$, 
with the corner parameters given by the labelling of $E(\cD)$.

b) Conversely, given a solution $z: V(\cG)\mapsto{\Bbb C}$ of the discrete 
Toda type system (\ref{Toda type}) on $\cG$, there exists a one--parameter 
family of its extensions to a solution of the system (\ref{eq prelim}) on 
$\cD$.
\end{proposition}
{\bf Proof.} The first statement is already proved. As for the second one,
suppose we know the solution of the Toda type system on $V(\cG)$, i.e. in
all ``black'' vertices. Fixing the value of the field $z$ in one ``white'' 
vertex arbitrarily, we can extend it succesively to all ``white'' vertices
in virtue of the equation (\ref{eq prelim}). The consistency
of this procedure is assured by (\ref{eq 3-leg}). \qed

\begin{remark} {\rm Proposition \ref{quadgraph to Toda} implies that from a 
solution of the Toda type system on $\cG$ one gets a one--parameter family of 
solutions of the Toda type system on $\cG^*$. This may be regarded as a sort 
of the {\itbf B\"acklund transformation}, especially when the graphs $\cG$ and 
$\cG^*$ are isomorphic, which is the case, e.g., for the regular square 
lattice.}
\end{remark}

\setcounter{equation}{0}
\section{Cross--ratio system and additive rational Toda system}
\label{Sect cross-ratio A}

\subsection{Definition}
The cross-ratio system is one of the simplest and at the same time the most
fundamental and important integrable systems on quad-graphs. 
\begin{definition}
Given a labelling of $E(\cD)$, the {\itbf cross--ratio system} on $\cD$ is the 
collection of equations for the function $z:V(\cD)\mapsto\widehat{{\Bbb C}}$, 
one equation per face of $\cD$, which read, in notations as on 
Fig.\,\ref{diamond again} and $z_k=z(v_k)$:
\begin{equation}\label{cross-rat eq}
q(z_0,z_1,z_2,z_3)\stackrel{\rm def}{=}\frac{(z_0-z_1)(z_2-z_3)}
{(z_1-z_2)(z_3-z_0)}=\frac{\alpha_1}{\alpha_2}\;.
\end{equation}
\end{definition}
Several simple remarks are here in order. Although (\ref{cross-rat eq}) is a
well--defined equation for each face of $\cD$, the cross-ratio itself is not a
function on $F(\cD)$, since
\[
q(z_0,z_1,z_2,z_3)=\frac{1}{q(z_1,z_2,z_3.z_0)}.
\]
On the other hand, $F(\cD)$ is in a one-to-one correspondence with $E(\cG)$,
as well as with $E(\cG^*)$. It is sometimes convenient to consider the
cross-ratio as a function on $E(\cG)$ or on $E(\cG^*)$, setting (see 
Fig.\,\ref{diamond again})
\begin{eqnarray}
q(\ee)=q(z_0,z_1,z_2,z_3),   & {\rm where} & \ee=(v_0,v_2)\in E(\cG),\\
q(\ee^*)=q(z_1,z_2,z_3.z_0), & {\rm where} & \ee^*=(v_1,v_3)\in E(\cG^*).
\end{eqnarray}

Sometimes a sort of inverse problem is of interest: given a function
$z:V(\cD)\mapsto\widehat{\Bbb C}$, can it be considered as a solution
of the cross--ratio system for {\it some} labelling? The answer follows 
from a simple lemma.
\begin{lemma}\label{lemma on cross-ratios}
Let, as usual, $\cD$ be a cellular decomposition of a topological disc
(open or closed), and therefore a double for some pair of dual
cellular decompositions $\cG$, $\cG^*$. Let $Q:E(\cG)\mapsto\widehat{\Bbb C}$ 
be some function, and extend it to $E(\cG^*)$ by the formula 
$Q(\ee^*)=1/Q(\ee)$. Then the necessary and sufficient condition for the 
existence of a labelling $\alpha: E(\cD)\mapsto{\Bbb C}$ such that, in
the notations of Fig.\,\ref{diamond again},
\begin{equation}\label{lemma 0}
Q(\ee)=Q(v_0,v_2)=\frac{\alpha_1}{\alpha_2}\quad\Leftrightarrow\quad
Q(\ee^*)=Q(v_1,v_3)=\frac{\alpha_2}{\alpha_1},
\end{equation}
is given (in the notations of Fig.\,\ref{star}) by the equations
\begin{equation}\label{lemma 1}
\prod_{\ee\in{\rm star}(v_0)} Q(\ee)=1\quad{\rm for}\;\;{\rm all}\;\;
{\rm internal}\;\; v_0\in V(\cG),
\end{equation}
and
\begin{equation}\label{lemma 2}
\prod_{\ee^*\in{\rm star}(v_1)} Q(\ee^*)=1\quad {\rm for}\;\;{\rm all}\;\;
{\rm internal}\;\; v_1\in V(\cG^*)
\end{equation}
\end{lemma}
{\bf Proof.} The necessity is obvious. To prove sufficiency, we construct 
$\alpha$ by assigning an arbitrary value (say, $\alpha=1$) to some edge from
$E(\cD)$, and then extending it successively using either of the 
equations (\ref{lemma 0}) and the definition of labelling. The
conditions (\ref{lemma 1}) and (\ref{lemma 2}) assure the consistency of this
procedure. \qed
\begin{corollary}\label{corollary on cross-ratios}
Let $z:V(\cD)\mapsto\widehat{\Bbb C}$ be some map. The ncessary and sufficient 
condition for the existence of a labelling $\alpha: E(\cD)\mapsto{\Bbb C}$ such 
that $z$ is a solution of the corresponding cross-ratio system is given (in the
notations of Fig.\,\ref{flower}) by the equation
\[
\prod_{k=1}^n q(z_0,z_{2k-1},z_{2k},z_{2k+1})=1
\]
for all internal vertices $v_0\in V(\cD)$.
\end{corollary}

\subsection{Zero curvature representation}

\begin{proposition}\label{cross-ratio Lax}
The cross--ratio system on $\cD$ admits a zero curvature representation
in the sense of Sect. \ref{Sect integrable systems on graphs}, with the
transition matrices for $\ee=(v_0,v_1)$: 
\begin{equation}\label{A Lax}
L(\ee,\lambda)=L(z_1,z_0,\alpha,\lambda)=
(1-\lambda\alpha)^{-1/2}\left(\begin{array}{cc}
1 & z_0-z_1 \\ \\ \displaystyle\frac{\lambda\alpha}{z_0-z_1} & 1\end{array}
\right).
\end{equation}
Here $\alpha=\alpha(\ee)$, and, as usual, $z_k=z(v_k)$.
\end{proposition}
{\bf Proof.} Verifying (\ref{zero curv inv}) and (\ref{quad-gr Lax}) is
a matter of a simple computation. Notice that the entry 12 
of the matrix equation (\ref{quad-gr Lax}) gives an identity, 
the entries 11 and 22 give the equation 
(\ref{cross-rat eq}) as it stands, while the entry 21 gives:
\begin{equation}\label{cross-rat aux}
\frac{\alpha_1}{z_0-z_1}+\frac{\alpha_2}{z_1-z_2}=
\frac{\alpha_2}{z_0-z_3}+\frac{\alpha_1}{z_3-z_2},
\end{equation}
which is a consequence (\ref{cross-rat eq}). \qed 
\vspace{2mm}

The zero--curvature representation of Proposition \ref{cross-ratio Lax}
was introduced in \cite{NC}, of course only in the case of the square lattice 
and constant labelling. Cf. also a gauge equivalent form in \cite{BP1}, where 
a more general situation of discrete isothermic surfaces in ${\Bbb R}^3$
is considered (in this context the fields $z$ take values in ${\Bbb H}$ -- 
the ring of quaternions).

\subsection{Zero curvature representation from three--dimensional consistency}

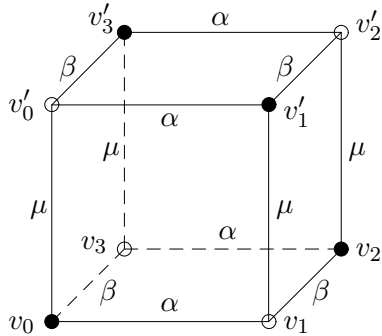
\begin{figure}[htbp]
\begin{center}
\setlength{\unitlength}{0.05em}
\begin{picture}(200,220)(0,0)
 \put(0,0){\circle*{10}}    \put(150,0){\circle{10}}
 \put(0,150){\circle{10}}   \put(150,150){\circle*{10}}
 \put(50,200){\circle*{10}} \put(200,200){\circle{10}}
 \put(50,50){\circle{10}}   \put(200,50){\circle*{10}}
 \path(0,0)(150,0)       \path(0,0)(0,150)
 \path(150,0)(150,150)   \path(0,150)(150,150)
 \path(0,150)(50,200)    \path(150,150)(200,200)   \path(50,200)(200,200)
 \path(200,200)(200,50) \path(200,50)(150,0) 
 \dashline[+30]{10}(0,0)(50,50)
 \dashline[+30]{10}(50,50)(50,200)
 \dashline[+30]{10}(50,50)(200,50)
 \put(-30,-5){$v_0$} \put(-30,145){$v_0'$}
 \put(160,-5){$v_1$} \put(160,140){$v_1'$}
 \put(210,45){$v_2$} \put(210,200){$v_2'$}
 \put(20,50){$v_3$}  \put(25,205){$v_3'$}
 \put(75,7){$\alpha$} \put(33,15){$\beta$}
 \put(115,57){$\alpha$} \put(180,15){$\beta$}
 \put(5,170){$\beta$} \put(75,135){$\alpha$}
 \put(155,170){$\beta$} \put(110,205){$\alpha$}
 \put(-15,75){$\mu$}\put(155,75){$\mu$}
 \put(35,115){$\mu$}\put(205,115){$\mu$}
\end{picture}
\caption{Elementary cube of $\cD\cup\cD'$}\label{cube}
\end{center}
\end{figure}

Now we demostrate how to {\it derive} the zero curvature representation of
Proposition \ref{cross-ratio Lax} starting from an additional deep property
of the cross--ratio system. To this end we extend the planar quad--graph $\cD$ 
into the third dimension. Formally speaking, we consider the second copy $\cD'$ 
of $\cD$ and add edges connecting each vertex $v\in V(\cD)$ with its copy 
$v'\in V(\cD')$. On this way we obtain a ``three--dimensional quad--graph'' 
${\bf D}$, whose set of vertices is 
\[
V({\bf D})=V(\cD)\cup V(\cD'),
\] 
whose set of edges is 
\[
E({\bf D})=E(\cD)\cup E(\cD')\cup\{(v,v'):v\in V(\cD)\},
\] 
and whose set of faces is 
\[
F({\bf D})=F(\cD)\cup F(\cD')\cup\{(v_0,v_1,v_1',v_0'):v_0,v_1\in V(\cD)\}.
\]
Elementary building blocks of $\bf D$ are ``cubes''
$(v_0,v_1,v_2,v_3,v_0',v_1',v_2',v_3')$, as shown on Fig.\,\ref{cube}.
Clearly, we still can consistently subdivide the vertices of $\bf D$ into
``black'' and ``white'' one, assigning to each $v'\in V(\cD')$ the colour
opposite to the colour of its counterpart $v\in V(\cD)$.

Next, we define a labelling on $E(\bf D)$ in the following way: each edge
$(v_0',v_1')\in E(\cD')$ carries the same label as its counterpart
$(v_0,v_1)\in E(\cD)$, while all ``vertical'' edges $(v,v')$ carry one and the
same label $\mu$. Now, the fundamental property of the cross--ratio system
mentioned above is the {\itbf three--dimensional consistency}. This should be 
understood as follows. Consider an elementary cube of $\bf D$, as on 
Fig.\,\ref{cube}. Suppose that the values of the field $z$ are 
given at the vertex $v_0$ and at its three neighbors $v_1$, $v_3$, and $v_0'$.
Then the cross--ratio equation (\ref{cross-rat eq}) uniquely determines the
values of $z$ at $v_2$, $v_1'$, and $v_3'$. After that the cross--ratio 
equation delivers three {\it \'a priori} different values for the value 
of the field $z$ at the vertex $v_2'$, coming from the faces
$(v_1,v_2,v_2',v_1')$, $(v_2,v_3,v_3',v_2')$, and $(v_0',v_1',v_2',v_3')$,
respectively. The three--dimensional consistency means
exactly that {\it these three values for $z(v_2')$ actually coincide}.
The verification of this claim consists of a straightforward computation.
Because of the importance of this property we formulate it in a separate
proposition. 
\begin{proposition}\label{3-dim compatibility} 
The cross--ratio system is consistent on the three--dimensional quad--graph 
${\bf D}$.
\end{proposition}

Now we can derive a zero curvature representation for the cross--ratio equation
on $\cD$, using the arbitraryness of the label $\mu$ assigned to the
``vertical'' edges $(v,v')$ of $\bf D$. For this aim, consider the cross--ratio
equation on the ``vertical'' face $(v_0,v_1,v_1',v_0')$. Denote
$\lambda=\mu^{-1}$, then the equation reads:
\begin{equation}\label{dSKdV Back}
\frac{(z_1'-z_0')(z_0-z_1)}{(z_0'-z_0)(z_1-z_1')}=\lambda\alpha.
\end{equation}
This gives $z_1'$ as a fractional--linear (M\"obius) transformation\footnote
{We use the following matrix notation for the action of M\"obius
transformations:
\[
\frac{az+b}{cz+d}=L[z],\quad{\rm where}\quad 
L=\left(\begin{array}{cc} a & b \\ c & d \end{array}\right).
\]
} 
of $z_0'$ with the coefficients depending on $z_0$, $z_1$, and the (arbitrary) 
parameter $\lambda$:
\[
z_1'=\widetilde{L}(z_1,z_0,\alpha,\lambda)[z_0'],
\]
where
\begin{eqnarray}
\lefteqn{\widetilde{L}(z_1,z_0,\alpha,\lambda)=
\left(\begin{array}{cc} 1+\displaystyle\frac{\lambda\alpha z_1}{z_0-z_1} & 
-\displaystyle\frac{\lambda\alpha z_0z_1}{z_0-z_1} \\ \\
\displaystyle\frac{\lambda\alpha}{z_0-z_1} &
1-\displaystyle\frac{\lambda\alpha z_0}{z_0-z_1}\end{array}\right)}\\
\nonumber\\
 & = & I+\frac{\lambda\alpha}{z_0-z_1}\left(\begin{array}{c} 
z_1 \\ 1 \end{array}\right)\left(\begin{array}{c} 
1 \\ -z_0\end{array}\right)^{\rm T} =
 (1-\lambda\alpha)I+\frac{\lambda\alpha}{z_0-z_1}\left(\begin{array}{c} 
z_0 \\ 1 \end{array}\right)\left(\begin{array}{c} 
1 \\ -z_1\end{array}\right)^{\rm T}.\quad \label{A Lax tilde} 
\end{eqnarray}
In principle, the matrices $\widetilde{L}(z_1,z_0,\alpha,\lambda)$, after 
multiplying by $(1-\lambda\alpha)^{-1/2}$ (which is nothing but normalizing 
the determinants) are the sought after transition matrices of the discrete 
flat connection on $\cD$, as assured by Proposition \ref{3-dim compatibility}. 
To recover the transition matrices from Proposition \ref{cross-ratio Lax},
we perform a gauge transformation defined on the level of the wave functions 
$\Psi:V(\cD)\mapsto G$ as
\begin{equation}\label{gauge Psi}
\Psi(v)\mapsto A(z(v))\Psi(v),
\end{equation}
where
\begin{equation}\label{A gauge}
A(z)=\left(\begin{array}{cc} 1 & z \\ \\ 0 & 1 \end{array}\right).
\end{equation}
On the level of transition matrices this results in
\[
\widetilde{L}(z_1,z_0,\alpha,\lambda)\mapsto L(z_1,z_0,\alpha,\lambda)=
A^{-1}(z_1)\widetilde{L}(z_1,z_0,\alpha,\lambda)A(z_0)=
\left(\begin{array}{cc} 1 & z_0-z_1 \\ \\
\displaystyle\frac{\lambda\alpha}{z_0-z_1} & 1 \end{array}\right).
\]  
Notice that the gauge transformation (\ref{gauge Psi}), (\ref{A gauge}) may be 
interpreted in terms of the variables $z(v')$ as a shift of each $z(v')$ by 
$z(v)$, so that the interpretation of the matrix $L(z_1,z_0,\alpha,\lambda)$ 
is as follows:
\[
z_1'-z_1=L(z_1,z_0,\alpha,\lambda)[z_0'-z_0].
\]

From our point of view, the property of the three--dimensional consistency 
lies at the heart of the two--dimensional integrability. Not only does it
allow us to derive the spectral--parameter dependent zero--curvature
representation for the two--dimensional discrete system. We can consider
the function $z:V(\cD')\mapsto\widehat{\Bbb C}$ as a 
{\itbf B\"acklund transformation} of the solution 
$z:V(\cD)\mapsto\widehat{\Bbb C}$. Moreover, if $\cD$ is generated by a square
lattice, then, as well--known, by refining this lattice in one or two 
directions, we can obtain certain integrable differential--difference or 
partial differential equations, the latter being the famous
Schwarzian Korteweg--de Vries equation
\[
z_t=z_{xxx}-\frac{3}{2}\frac{z_{xx}^2}{z_x},
\] 
see \cite{NC}. So, the three--dimensional consistency delivers all these
equations and their B\"acklund transformations in one construction.
For instance, if in (\ref{dSKdV Back}) $z_0$ is a solution of the
Schwarzian KdV, $z_1$ its B\"acklund transformation, and $z_0',z_1'$
correspond to the $\epsilon$--shift of these solutions in the $x$--direction, 
where $\lambda=\epsilon^{-2}$, then in the limit $\epsilon\to 0$ one recovers
the well--known formula \cite{A1}
\[
(z_0)_x(z_1)_x=\alpha^{-1}(z_0-z_1)^2.
\]
Similarly, when considered on a single face of the diamond, as on 
Fig.\,\ref{diamond again}, the cross-ratio equation (\ref{cross-rat eq})
expresses nothing but the commutativity of the Bianchi's 
diagram (Fig.\,\ref{diamond again}) for the B\"acklund transformations of the 
Schwarzian KdV. Indeed, multiplying the equations 
\[
(z_0)_x(z_1)_x=\alpha_1^{-1}(z_0-z_1)^2,\qquad 
(z_3)_x(z_2)_x=\alpha_1^{-1}(z_3-z_2)^2,
\]
and dividing by two similar equations 
corresponding to the edges $(v_0,v_3)$ and $(v_1,v_2)$ (with the parameter 
$\alpha_2$), we arrive at the (squared) equation (\ref{cross-rat eq}). So,
one can say that the equation (\ref{cross-rat eq}) on the cube Fig.\,\ref{cube}
contains {\it everything} about the Schwarzian KdV and its B\"acklund
transformations.

\subsection{Additive rational Toda system}

The further remarkable feature of the cross--ratio equation is the existence
of the {\itbf three-leg form}. To derive it, notice that (\ref{cross-rat eq})
is equivalent to
\[
\alpha_1\,\frac{z_1-z_2}{z_0-z_1}-\alpha_2\,\frac{z_3-z_2}{z_0-z_3}=0\quad
\Leftrightarrow\quad
\alpha_1\,\frac{z_0-z_2}{z_0-z_1}-\alpha_2\,\frac{z_0-z_2}{z_0-z_3}=
\alpha_1-\alpha_2,
\]
or to 
\begin{equation}\label{cross-rat 3-leg}
\frac{\alpha_1}{z_0-z_1}-\frac{\alpha_2}{z_0-z_3}=
\frac{\alpha_1-\alpha_2}{z_0-z_2},
\end{equation}
which obviously is of the form (\ref{eq 3-leg}).

According to the construction of Sect.\,\ref{Sect from diamond to Toda}, 
the corresponding system of the Toda type reads (in the notations of 
Fig.\,\ref{star}):
\begin{equation}\label{Toda A}
\sum_{k=1}^n\frac{\alpha_k-\alpha_{k+1}}{z_0-z_{2k}}=0.
\end{equation}
It will be called the {\itbf additive rational Toda system} on $\cG$. 
For the cross--ratio system on $\cD$ and additive rational Toda systems
on $\cG$, $\cG^*$ there holds Proposition \ref{quadgraph to Toda}.

\begin{remark} {\rm Notice that the Toda system on $\cG$ remains unchanged
upon the {\it simultaneous additive shift} of all corner parameters
$\alpha_k\mapsto\alpha_k+a$. However, such a shift changes the cross--ratio 
system essentially. Hence, Proposition \ref{quadgraph to Toda} yields also a 
nontrivial transformation between the solutions of two different cross--ratio
systems on $\cD$ whose labellings differ by a common additive shift.}
\end{remark}

Next, we derive a zero curvature representation for the additive rational Toda
system from such a representation for the cross--ratio system. Consider the 
additive rational Toda system on the ``black'' vertices from $V(\cG)$, i.e. on 
$v_{2k}$ in the notations of Fig.\,\ref{flower}. The transition matrix across 
the edge $\ee_k=(v_0,v_{2k})\in E(\cG)$, i.e. along the edge
$\ee_k^*=(v_{2k-1},v_{2k+1})\in E(\cG^*)$, is given by
\begin{equation}\label{L cross-rat to Toda A}
\widetilde{\cL}(z_{2k},z_0,\lambda)=
\widetilde{L}(z_{2k+1},z_0,\alpha_{k+1},\lambda)
\widetilde{L}(z_0,z_{2k-1},\alpha_k,\lambda).
\end{equation}
We use here the matrices (\ref{A Lax tilde}), gauge equivalent to (\ref{A Lax}),
because of the remarkable fact underlined by the notation for the matrices on
the left--hand side of the last equation: they actually depend only on $z_0,
z_{2k}$, but not on $z_{2k-1},z_{2k+1}$. (This propery would fail, if we would 
use the matrices (\ref{A Lax}).) 
\begin{proposition}\label{Lax cross-rat to Toda A}
\begin{eqnarray}
\lefteqn{\widetilde{\cL}(z_{2k},z_0,\lambda)=
I+\frac{\lambda}{z_0-z_{2k}}
\left(\begin{array}{cc} \alpha_{k+1}z_{2k}-\alpha_kz_0 &
(\alpha_k-\alpha_{k+1})z_0z_{2k} \\ \\ \alpha_{k+1}-\alpha_k & 
\alpha_kz_{2k}-\alpha_{k+1}z_0\end{array}\right)}\label{L Toda A}\\ 
\nonumber\\
& = & (1-\lambda\alpha_k)I+
\lambda\frac{\alpha_{k+1}-\alpha_k}{z_0-z_{2k}}
\left(\begin{array}{c} z_{2k} \\ 1 \end{array}\right)
\left(\begin{array}{c} 1 \\ -z_0   \end{array}\right)^{\rm T}
\label{L Toda A rank 1 1}\\
 & = & (1-\lambda\alpha_{k+1})I+\lambda\frac{\alpha_{k+1}-\alpha_k}{z_0-z_{2k}}
\left(\begin{array}{c} z_0 \\ 1 \end{array}\right)
\left(\begin{array}{c} 1 \\ -z_{2k}  \end{array}\right)^{\rm T}.
\label{L Toda A rank 1 2}
\qquad \qquad
\end{eqnarray}
\end{proposition}
{\bf Proof.} Use the second formula in (\ref{A Lax tilde}) for 
$\widetilde{L}(z_0,z_{2k-1},\alpha_k,\lambda)$
and the first formula in (\ref{A Lax tilde}) for
$\widetilde{L}(z_{2k+1},z_0,\alpha_{k+1},\lambda)$. 
This leads, after a remarkable
cancellation of terms quadratic in $\lambda$, to the following expression
for the matrix (\ref{L cross-rat to Toda A}):
\[
\widetilde{\cL}(z_{2k},z_0,\lambda)=(1-\lambda\alpha_k)I+
\lambda \xi\left(\begin{array}{c} 1 \\ -z_0 \end{array}\right)^{\rm T},
\]
where
\[
\xi=\frac{\alpha_{k+1}}{z_0-z_{2k+1}}\left(\begin{array}{c} z_{2k+1} \\ 1
\end{array}\right)-
\frac{\alpha_k}{z_0-z_{2k-1}}\left(\begin{array}{c} z_{2k-1} \\ 1
\end{array}\right).
\]
A simple calculation based on the formula (\ref{cross-rat 3-leg}), written 
now as
\[
\frac{\alpha_{k+1}}{z_0-z_{2k+1}}-\frac{\alpha_k}{z_0-z_{2k-1}}=
\frac{\alpha_{k+1}-\alpha_k}{z_0-z_{2k}},
\]
shows that 
\[
\xi=\frac{\alpha_{k+1}-\alpha_k}{z_0-z_{2k}}\left(\begin{array}{c} 
z_{2k} \\ 1 \end{array}\right).
\]
This proves (\ref{L Toda A rank 1 1}).  \qed
\vspace{2mm}

The matrix (\ref{L Toda A}) coincides with the transition matrix 
for the additive rational Toda lattice found by V.Adler in \cite{A3}.

\subsection{Duality}

Next, we discuss a duality transformation for the cross--ratio system. 
This comes into play, if we rewrite the cross--ratio equation
(\ref{cross-rat aux}) in the form
\begin{equation}\label{cross-rat dual}
\frac{\alpha_1}{z_0-z_1}+\frac{\alpha_2}{z_1-z_2}+\frac{\alpha_1}{z_2-z_3}
+\frac{\alpha_2}{z_3-z_0}=0.
\end{equation}
This guarantees (locally, and therefore in the case when $\cD$ is a
cellular decomposition of a topological disc) that one can introduce 
the function $Z:V(\cD)\mapsto\widehat{\Bbb C}$ such that
\begin{equation}\label{cr dual}
Z_1-Z_0=\frac{\alpha_1}{z_0-z_1},\quad Z_2-Z_1=\frac{\alpha_2}{z_1-z_2},\quad
Z_3-Z_2=\frac{\alpha_1}{z_2-z_3},\quad Z_0-Z_3=\frac{\alpha_2}{z_3-z_0}.
\end{equation}
In other words, for an arbitrary $\ee=(v_1,v_2)\in E(\cD)$, one sets
\begin{equation}\label{A dual edge}
Z(v_2)-Z(v_1)=\frac{\alpha(\ee)}{z(v_1)-z(v_2)}.
\end{equation}
\begin{proposition}\label{A dual}
For any solution $z:V(\cD)\mapsto\widehat{\Bbb C}$ of the cross--ratio system on
$\cD$, the formula (\ref{A dual edge}) correctly defines a new (dual) solution
of the same system. 
\end{proposition}
{\bf Proof.} The above argument shows that the formula (\ref{A dual edge}) 
correctly defines some function on $V(\cD)$. It can be easily seen that this
function also solves the cross--ratio system. \qed
\vspace{2mm}

Combining Propositions \ref{quadgraph to Toda} and \ref{A dual}, we see that
to any solution of the additive rational Toda system on $\cG$ there corresponds
a one--parameter family of solutions of the additive rational Toda system on 
$\cG^*$, and any such pair of corresponding solutions yields (almost uniquely 
-- up to an additive constant) dual solutions for the additive rational Toda 
systems on $\cG$ and $\cG^*$. But actually this duality relation between Toda 
systems is much more stiff. Indeed, adding 
the first and the fourth equations in (\ref{cr dual}) and using 
(\ref{cross-rat 3-leg}), we arrive at the formula 
\begin{equation}\label{Toda A dual}
Z_1-Z_3=\frac{\alpha_2-\alpha_1}{z_2-z_0}.
\end{equation}
Therefore, a solution of the additive rational Toda system on $\cG$ determines 
the dual solution of the additive rational Toda system on $\cG^*$ almost 
uniquely (up to an additive constant).
\vspace{2mm}

It is also instructive to regard the function $w: V(\cD)\mapsto\widehat{\Bbb C}$
which coincides with $z$ on $V(\cG)$ and with $Z$ on $V(\cG^*)$. Then 
formula (\ref{Toda A dual}) may be considered as a system of equations on 
the quad-graph, reading on the face $(v_0,v_1,v_2,v_3)$ as
\begin{equation}\label{KdV NSP}
w_1-w_3=\frac{\alpha_2-\alpha_1}{w_2-w_0}.
\end{equation}
Interestingly (but not very surprizingly), the equations (\ref{KdV NSP}) also 
constitute an integrable system on the quad-graph. In the case when $\cD$ is
generated by a square lattice and the labelling is constant, this system is
known under the name of the ``discrete KdV equation'' \cite{H1}, \cite{NC}.
Actually, it appeared even earlier in the numerical analysis as the
``$\varepsilon$--algorithm'' \cite{W}.  The transition matrices for the 
equation (\ref{KdV NSP}) are given by 
\begin{equation}\label{KdV NSP Lax}
L(\ee,\mu)=L(w_1,w_0,\alpha,\mu)=
(\mu-\alpha)^{-1/2}\left(\begin{array}{cc}
w_0 & -w_0w_1+\alpha-\mu \\ \\ 1 & -w_1 \end{array}\right).
\end{equation}
This transition matrix may be derived in exactly the same way as for the
cross--ratio system, based on the three--dimensional consistency.
Namely, this is the matrix of the M\"obius transformation
$v_0'\mapsto v_1'$, see Fig.\,\ref{cube}:
\[
(w_1-w_0')(w_1'-w_0)+\alpha-\mu=0\quad\Rightarrow\quad
w_1'=\frac{w_0w_0'-w_0w_1+\alpha-\mu}{w_0'-w_1}=
L(w_1,w_0,\alpha,\mu)[w_0'].
\]
Again, the three--dimensional consistency is the central principle for
the whole integrability theory of the discrete, differential--difference and
partial differential equations related to the 
(potential) Korteweg--de Vries equation
\[
w_t=w_{xxx}-6w_x^2,
\]
as well as their B\"acklund transformations.
In particular, the equation (\ref{KdV NSP}) expresses nothing but the 
commutativity of the Bianchi's diagram (Fig.\,\ref{diamond again}) for the 
B\"acklund transformations of the potential KdV. Indeed, recall that the 
B\"acklund transformations corresponding to the edges
$(v_0,v_1)$ and $(v_3,v_2)$ of the diagram Fig.\,\ref{diamond again} are 
defined by the equations
\[
(w_0)_x+(w_1)_x=(w_1-w_0)^2+\alpha_1,\qquad 
(w_3)_x+(w_2)_x=(w_2-w_3)^2+\alpha_1.
\]
Adding these two equations and subtracting two similar equations corresponding
to the edges $(v_0,v_3)$ and $(v_1,v_2)$ (with the parameter $\alpha_2$), we 
arrive at
\[
(w_2-w_0)(w_1-w_3)+\alpha_1-\alpha_2=0,
\]
which is nothing but (\ref{KdV NSP}).
\vspace{2mm}

Notice that equation (\ref{KdV NSP}) is already in the three--leg form, 
and restrictions of its solutions to $V(\cG)$ and to $V(\cG^*)$ are solutions 
of the rational additive Toda system on the corresponding graphs.
It is easy to check that the pairwise products of the matrices 
(\ref{KdV NSP Lax}), like in (\ref{L cross-rat to Toda A}), lead to the
same transition matrices (\ref{L Toda A}) for the additive rational Toda 
system as before.

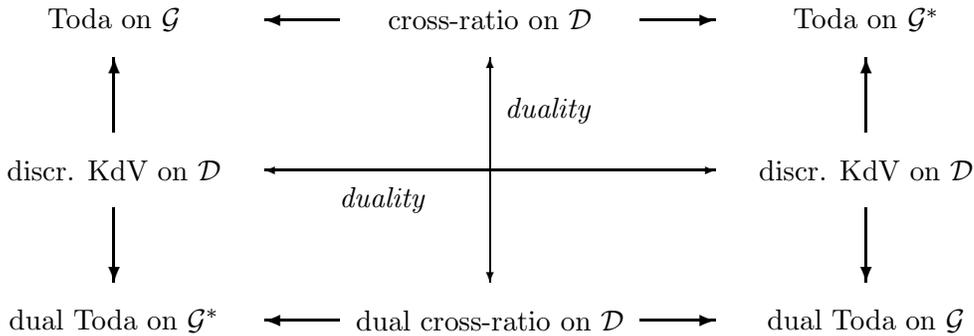
\begin{figure}[htbp]
\begin{center}
\unitlength1cm
\begin{picture}(16,5)
\put(7.5,2.5){\vector(1,0){3}}
\put(7.5,2.5){\vector(-1,0){3}}
\put(7.5,2.5){\vector(0,1){1.5}}
\put(7.5,2.5){\vector(0,-1){1.5}}
\put(7.7,3.2){{\it duality}}
\put(5.5,2){{\it duality}}
\thicklines
\put(1,4){\makebox(3,1){Toda on $\cG$}}
\put(6,4){\makebox(3,1){cross-ratio on $\cD$}}
\put(11,4){\makebox(3,1){Toda on $\cG^*$}}
\put(5.5,4.5){\vector(-1,0){1}}
\put(9.5,4.5){\vector(1,0){1}}
\put(1,0){\makebox(3,1){dual Toda on $\cG^*$}}
\put(6,0){\makebox(3,1){dual cross-ratio on $\cD$}}
\put(11,0){\makebox(3,1){dual Toda on $\cG$}}
\put(5.5,0.5){\vector(-1,0){1}}
\put(9.5,0.5){\vector(1,0){1}}
\put(1,2){\makebox(3,1){discr. KdV on $\cD$}}
\put(11,2){\makebox(3,1){discr. KdV on $\cD$}}
\put(2.5,3){\vector(0,1){1}}
\put(12.5,3){\vector(0,1){1}}
\put(2.5,2){\vector(0,-1){1}}
\put(12.5,2){\vector(0,-1){1}}
\end{picture}
\caption{Relations between different systems}\label{diagram}
\end{center}
\end{figure}

The relations between various systems discussed in the present section are
summarized on Fig.\,\ref{diagram}. 

The thick lines here are for the
restriction maps, the thin ones  -- for the duality maps. The duality
for the cross-ratio system is described by the formula (\ref{A dual edge}).
Its transcription in terms of the variables $w$ and the dual ones $W$ reads:
\begin{equation}\label{dKdV dual edge}
W(v_2)-w(v_1)=\frac{\alpha(\ee)}{W(v_1)-w(v_2)}.
\end{equation}
It is not difficult to prove the statement similar to Proposition \ref{A dual}:
For any solution $w:V(\cD)\mapsto\widehat{\Bbb C}$ of the discrete KdV system 
on $\cD$, the formula (\ref{dKdV dual edge}) correctly defines a one--parameter
family of (dual) solutions $W:V(\cD)\mapsto\widehat{\Bbb C}$ of the same 
system.

\subsection{Cross--ratio system and circle patterns}
\label{subsect patterns A}
We close the discussion of the cross--ratio system on an arbitrary quad-graph 
$\cD$ by giving its geometric interpretation. For this aim, we consider
circle patterns with arbitrary combinatorics.
\begin{definition}
Let $\cG$ be an arbitrary cellular decomposition. We say that a map
$z:V(\cG)\mapsto \widehat{{\Bbb C}}$ defines a {\itbf circle pattern with
the combinatorics of} $\cG$, if the following condition is satisfied.
Let $f\in F(\cG)$ be an arbitrary face of $\cG$, and let $v_1,v_2,\ldots,v_n$
be its consecutive vertices. Then the points $z(v_1),z(v_2),\ldots,z(v_n)\in
\widehat{{\Bbb C}}$ lie on a circle, and their circular order is just the 
listed one. We denote this circle by $C(f)$, thus putting it into a 
correspondence with the face $f$, or, equivalently, with the respective 
vertex of the dual decomposition $\cG^*$.
\end{definition}
As a consequence of this condition, if two faces $f_1,f_2\in F(\cG)$ have a
common edge $(v_1,v_2)$, then the circles $C(f_1)$ and $C(f_2)$ intersect in
the points $z(v_1),z(v_2)$. In other words, the edges from $E(\cG)$ correspond
to pairs of neighboring (intersecting) circles of the pattern.
Similarly, if several faces $f_1,f_2,\ldots,f_m\in F(\cG)$ meet in one 
point $v_0\in V(\cG)$, then the corresponding circles
$C(f_1),C(f_2),\ldots,C(f_m)$ also have a common intersection point $z(v_0)$.

\begin{figure}[htbp]
\begin{center}
\setlength{\unitlength}{0.025em}
\begin{picture}(1000,800)(-700,-400)
 \put(0,0){\circle{20}}
 \put(0,0){\circle{240}} 
 \put(5,119.8958){\circle*{20}}
 \put(118.3216,20){\circle*{20}}
 \put(90,-79.3725){\circle*{20}}
 \put(-90,-79.3725){\circle*{20}}
 \put(-90,79.3725){\circle*{20}}
 \put(0,-190){\circle{20}}
 \put(0,-190){\circle{285.2258}} 
 \put(208.3216,-59.3725){\circle{20}}
 \put(208.3216,-59.3725){\circle{240}}
 \put(-370,0){\circle{20}}
 \put(-370,0){\circle{582.06527}}
 \put(-85,198.2683){\circle{20}}
 \put(-85,198.2683){\circle{240}}
 \put(123.3216,139.8958){\circle{20}}
 \put(123.3216,139.8958){\circle{240}}
 \path(5,119.8958)(118.3216,20)
 \path(118.3216,20)(90,-79.3725)
 \path(90,-79.3725)(-90,-79.3725)
 \path(-90,-79.3725)(-90,79.3725)
 \path(-90,79.3725)(5,119.8958)
 \dashline[+30]{10}(0,0)(0,-190)
 \dashline[+30]{10}(0,0)(208.3216,-59.3725)
 \dashline[+30]{10}(0,0)(-370,0)
 \dashline[+30]{10}(0,0)(-85,198.2683)
 \dashline[+30]{10}(0,0)(123.3216,139.8958)
 \thicklines
 \path(-370,0)(-90,-79.3725)  \path(-370,0)(-90,79.3725)
 \path(0,0)(-90,-79.3725)  \path(0,0)(-90,79.3725)
 \path(-85,198.2683)(-90,79.3725)  \path(-85,198.2683)(5,119.8958)
 \path(0,0)(5,119.8958)
 \path(123.3216,139.8958)(5,119.8958)  \path(123.3216,139.8958)(118.3216,20)  
 \path(0,0)(118.3216,20)
 \path(208.3216,-59.3725)(118.3216,20) \path(208.3216,-59.3725)(90,-79.3725)
 \path(0,0)(90,-79.3725)
 \path(0,-190)(90,-79.3725)  \path(0,-190)(-90,-79.3725)
    \end{picture}
\caption{Circle pattern}\label{circle pattern}
\end{center}
\end{figure}
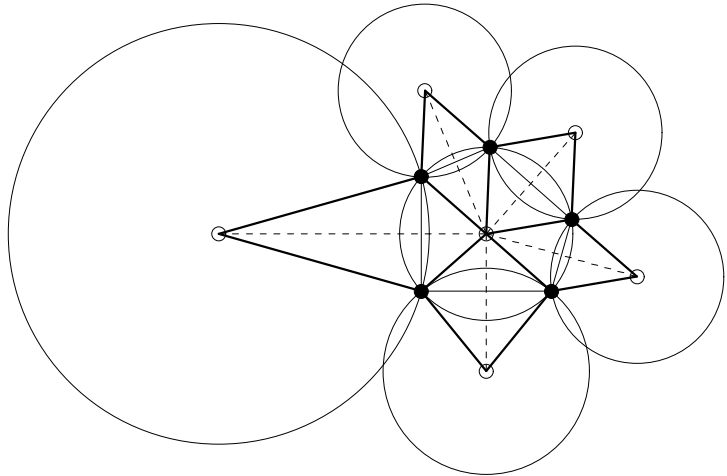

Given a circle pattern with the combinatorics of $\cG$, we can extend the
function $z$ to the vertices of the dual graph, simply setting
\[
z(f)={\rm Euclidean\;\;center\;\;of\;\;the\;\;circle}\;\;C(f),
\quad f\in F(\cG)\simeq V(\cG^*).
\]
A slightly more general extension is achieved by setting
\[
z(f)={\rm conformal\;\;center\;\;of\;\;the\;\;circle}\;\;C(f),
\quad f\in F(\cG)\simeq V(\cG^*).
\]
Recall that the conformal center of a circle $C$ is a reflection of some fixed
point $P_{\infty}\in\widehat{\Bbb C}$ in $C$. In particular, if $P_{\infty}=
\infty$, then the conformal center coincides with the Euclidean one.
 
Thus, after either extension, the map $z$ is defined on all of 
$V(\cD)=V(\cG)\cup V(\cG^*)$, where $\cD$ is the double of $\cG$. 
Consider a face of the double. Clearly, its vertices $v_0,v_1,v_2,v_3$ 
correspond to the intersection points and the (conformal) centers of two 
neighboring circles $C_0,C_1$ of the pattern. Let $z_0,z_2$ be the 
intersection points, and let $z_1,z_3$ be the centers of the circles. 
See Fig.\,\ref{two circles}.
\begin{figure}[htbp]
\begin{center}
\setlength{\unitlength}{0.04em}
\begin{picture}(500,320)(-300,-160)
 \put(-150,0){\circle{10}}
 \put(-155,-25){$z_3$}
 \put(50,0){\circle{10}} 
 \put(45,-25){$z_1$}
 \put(0,50){\circle*{10}}
 \put(-30,55){$z_0$}
 \put(0,-50){\circle*{10}}
 \put(-5,-75){$z_2$}
 \put(50,0){\circle{141.42136}}
 \put(130,0){$C_1$}
 \put(-150,0){\circle{316.22777}}
 \put(-295,0){$C_0$}
 \path(0,50)(0,-50)
 \dashline[+30]{9}(-150,0)(-110,0)
 \dashline[+30]{10}(-85,0)(50,0)
 \dottedline{6}(0,50)(50,100)
 \dottedline{6}(0,50)(-25,125)
 \thicklines
 \path(-150,0)(0,50)  \path(0,50)(50,0)
 \path(50,0)(0,-50)  \path(0,-50)(-150,0)
  \put(-1,68){$\phi$}
  \put(-105,-5){$\psi$}
    \end{picture}
\caption{Two intersecting circles}\label{two circles}
\end{center}
\end{figure}
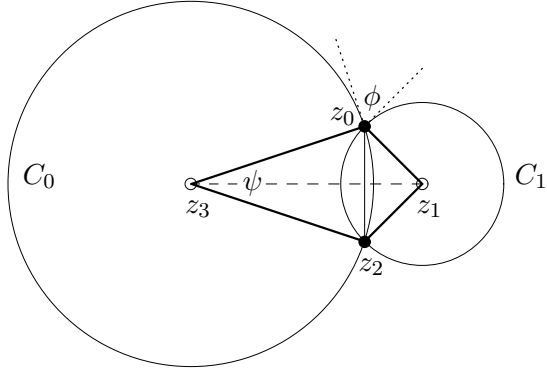
\begin{lemma}\label{lemma on angles}
If the intersection angle of $C_0,C_1$ is equal to $\phi$, then
\begin{equation}
q(z_0,z_1,z_2,z_3)=\exp(2i\phi).
\end{equation}
\end{lemma}
{\bf Proof.} The claim is obvious for the Euclidean centers. These can be 
mapped onto conformal centers by an appropriate M\"obius transformation
which preserves the cross--ratios. \qed
\begin{proposition}\label{labelling for pattern}
Let $\cG$ be a cellular decomposition of a topological disc (open or closed),
and consider a circle pattern with the combinatorics of $\cG$. Let 
$z:V(\cD)\mapsto\widehat{\Bbb C}$ come from a circle pattern,
i.e. $\{z(v):v\in V(\cG)\}$ consists of intersection points of the circles,
and $\{z(v):v\in V(\cG^*)\}$ consists of their conformal centers. Suppose 
that the intersection angles of the circles satisfy the following condition:
\begin{itemize}
\item For each circle of the pattern the sum of its intersection 
angles $\phi_k$, $k=1,2,\ldots,m$ with all neighboring circles
of the pattern satisfies $2\sum_{k=1}^m\phi_k\equiv 0\pmod{2\pi}$.
\end{itemize}
Then there exists a labelling $\alpha:E(\cD)\mapsto{\Bbb S}^1=
\{\theta\in{\Bbb C}:|\theta|=1\}$ depending only on the intersection angles,
such that $z$ is a solution of the corresponding cross--ratio system. 
\end{proposition}
{\bf Proof.} According to Corollary \ref{corollary on cross-ratios}, the 
necessary and sufficient condition for the claim of the proposition to hold 
is given by (\ref{lemma 1}), (\ref{lemma 2}). The second of these two formulas 
is exactly the condition formulated in the proposition. The first one is
equivalent to the following condition: for an arbitrary common intersection 
point of $n$ circles the sum of their pairwise intersection angles 
$\phi_1,\ldots,\phi_n$ satisfies the similar condition 
$2\sum_{j=1}^n\phi_j\equiv 0\pmod{2\pi}$. This condition 
is satisfied automatically for geometrical reasons (its left--hand 
side is equal to $2\pi$). \qed

\begin{corollary} Under the condition of the previous proposition, the 
restriction of the function $z$ to $V(\cG)$ (i.e. the intersection points
of the circles) satisfies the additive rational Toda system on $\cG$, while 
the restriction of the function $z$ to $V(\cG^*)$ (i.e. the conformal centers
of the circles) satisfies the type (A) Toda system on $\cG^*$.
\end{corollary}

\setcounter{equation}{0}
\section{Shifted ratio system and multiplicative rational Toda system}
\label{Sect cross-ratio C}

We now define a system on a quad-graph $\cD$ which can be considered as a
deformation of the cross--ratio system. 
\begin{definition}
Given a labelling of $E(\cD)$, the {\itbf shifted ratio system} on $\cD$ is
the collection of equations for the function $z:V(\cD)\mapsto{\Bbb C}$, one
equation per face of $\cD$, which read, in notations as on Fig. \ref{diamond
again} and $z_k=z(v_k)$:
\begin{equation}\label{shifted rat eq 1}
\frac{(z_0-z_1+\alpha_1)(z_2-z_3+\alpha_1)}
{(z_1-z_2-\alpha_2)(z_3-z_0-\alpha_2)}=\frac{\alpha_1}{\alpha_2}\;,
\end{equation}
or, equivalently (as one checks directly),
\begin{equation}\label{shifted rat eq 2}
\frac{(z_0-z_1-\alpha_1)(z_2-z_3-\alpha_1)}
{(z_1-z_2+\alpha_2)(z_3-z_0+\alpha_2)}=\frac{\alpha_1}{\alpha_2}\;.
\end{equation}
\end{definition}
The cross--ratio system of the previos section may be obtained from
the shifted ratio system by the following limit: replace in 
(\ref{shifted rat eq 1}) $\alpha_k$ by $\epsilon\alpha_k$ and then send 
$\epsilon\to 0$. The theory of the shifted ratio system is 
parallel to that of the cross--ratio system, therefore our presentation here
will be more tense.
\begin{proposition}\label{shifted ratio Lax}
The shifted ratio system on $\cD$ admits a zero curvature representation
in the sense of Sect. \ref{Sect integrable systems on graphs}, with the
transition matrices for $\ee=(v_0,v_1)$: 
\begin{equation}\label{C Lax}
L(\ee,\lambda)=L(z_1,z_0,\alpha,\lambda)=
(1-\lambda\alpha)^{-1/2}\left(\begin{array}{cc}
1 & z_0-z_1+\alpha \\ \\ \displaystyle\frac{\lambda\alpha}{z_0-z_1-\alpha} & 
\displaystyle\frac{z_0-z_1+\alpha}{z_0-z_1-\alpha}  \end{array}\right).
\end{equation}
Here $\alpha=\alpha(\ee)$, and $z_k=z(v_k)$.
\end{proposition}
{\bf Proof.} As in the proof of Proposition \ref{cross-ratio Lax}, we have 
to verify the equations (\ref{zero curv inv}) and (\ref{quad-gr Lax}). This
is done again by a direct computation. In particular, one checks that
the entry 11 and the $\lambda$--term of the entry 22 of the matrix equation 
\[
L(z_2,z_1,\alpha_2,\lambda)L(z_1,z_0,\alpha_1,\lambda)=
L(z_2,z_3,\alpha_1,\lambda)L(z_3,z_0,\alpha_2,\lambda).
\]
are equivalent to the equations (\ref{shifted rat eq 2}) and 
(\ref{shifted rat eq 1}), respectively.
The $\lambda^0$--term of the entry 22 reads:
\begin{equation}\label{shifted rat aux}
\frac{z_1-z_2+\alpha_2}{z_1-z_2-\alpha_2}\cdot
\frac{z_0-z_1+\alpha_1}{z_0-z_1-\alpha_1}=
\frac{z_3-z_2+\alpha_1}{z_3-z_2-\alpha_1}\cdot
\frac{z_0-z_3+\alpha_2}{z_0-z_3-\alpha_2},
\end{equation}
which is a plain consequence of (\ref{shifted rat eq 1}), 
(\ref{shifted rat eq 2}). Further, the entry 12 reads:
\[
z_0-z_1+\alpha_1+(z_1-z_2+\alpha_2)\,\frac{z_0-z_1+\alpha_1}{z_0-z_1-\alpha_1}=
z_0-z_3+\alpha_2+(z_3-z_2+\alpha_1)\,\frac{z_0-z_3+\alpha_2}{z_0-z_3-\alpha_2},
\] 
which is equivalent to
\begin{equation}\label{shifted rat 3-leg}
\frac{z_0-z_1+\alpha_1}{z_0-z_1-\alpha_1}\cdot
\frac{z_0-z_3-\alpha_2}{z_0-z_3+\alpha_2}=
\frac{z_0-z_2+\alpha_1-\alpha_2}{z_0-z_2-\alpha_1+\alpha_2}.
\end{equation}
The latter equation (after taking logarithm) is nothing but the 
{\itbf three--leg form} of the shifted ratio equation. Indeed, straightforward 
algebraic manipulations show that it is equivalent to (\ref{shifted rat eq 1}) 
(or to (\ref{shifted rat eq 2})).
Finally, consider the entry 21 of the above matrix equation:
\[
\frac{\alpha_2}{z_1-z_2-\alpha_2}+\frac{\alpha_1}{z_0-z_1-\alpha_1}\cdot
\frac{z_1-z_2+\alpha_2}{z_1-z_2-\alpha_2}=
\frac{\alpha_1}{z_3-z_2-\alpha_1}+\frac{\alpha_2}{z_0-z_3-\alpha_2}\cdot
\frac{z_3-z_2+\alpha_1}{z_3-z_2-\alpha_1}.
\]
With the help of (\ref{shifted rat eq 2}) this may be rewritten as
\[
\frac{\alpha_2}{z_1-z_2-\alpha_2}\left(1+
\frac{z_2-z_3-\alpha_1}{z_3-z_0+\alpha_2}\right)=
\frac{\alpha_1}{z_3-z_2-\alpha_1}\left(1+
\frac{z_1-z_2+\alpha_2}{z_0-z_1-\alpha_1}\right),
\] 
and this, in turn, is rewritten with the help of (\ref{shifted rat eq 1}) as
(\ref{shifted rat 3-leg}). \qed
\vspace{2mm}

The shifted ratio system, as well as its generalization from the next Section,
appeared in \cite{NQC} (of course, only in the simplest possible 
situation: regular square lattice, labelling constant along each of the two 
lattice directions). They denoted this equation as ``the discrete 
Krichever--Novikov system''. The zero curvature representation was given
later in \cite{NC}. Actually, this zero curvature representation can be
{\it derived} along the same lines as for the cross--ratio system, based
on the {\itbf three--dimensional consistency}, which remains to hold true 
also in the present situation. 
\begin{proposition}\label{shifted rat 3-dim compatibility} 
The shifted ratio system is consistent on the three--dimensional quad--graph 
${\bf D}$.
\end{proposition}
{\bf Proof} is again a matter of straightforward computations. \qed
\vspace{2mm}

To derive a zero curvature representation from this fact, we proceed as in the
previous section: assign the label $\mu$ to all ``vertical'' edges $(v,v')$ of 
$\bf D$, and consider the shifted ratio equation on the ``vertical'' face 
$(v_0,v_1,v_1',v_0')$:
\[
\frac{(z_1'-z_0'+\alpha)(z_0-z_1+\alpha)}
{(z_0'-z_0-\mu)(z_1-z_1'-\mu)}=\frac{\alpha}{\mu}.
\]
Denote $\lambda=\mu^{-1}$, and solve for $z_1'$ in terms of $z_0'$:
\[
z_1'=\widetilde{L}(z_1,z_0,\alpha,\lambda)[z_0'],
\]
where
\[
\widetilde{L}(z_1,z_0,\alpha,\lambda)=\left(\begin{array}{cc}
z_0-z_1+\lambda\alpha & \lambda^{-1}\alpha-\alpha^2-\lambda\alpha z_0z_1 \\ \\
\lambda\alpha & z_0-z_1-\lambda\alpha z_0\end{array}\right).
\]
(Recall that the limit to the formulas of the previous section is the following
one: $\alpha\mapsto\epsilon\alpha$, $\mu\mapsto\epsilon\mu$, so that 
$\lambda\mapsto \epsilon^{-1}\lambda$, and then $\epsilon\to 0$.)
To recover the matrices of Proposition \ref{shifted ratio Lax}, perform the
gauge transformation 
\begin{equation}\label{gauge Psi C}
\Psi(v)\mapsto C(z(v),\lambda)\Psi(v),
\end{equation}
where
\begin{equation}\label{C gauge}
C(z,\lambda)=\left(\begin{array}{cc} 1 & z-\lambda^{-1} \\ \\ 0 & 1 
\end{array}\right)=\left(\begin{array}{cc} 1 & -\lambda^{-1} \\ \\ 0 & 1 
\end{array}\right)A(z).
\end{equation}
An easy computation shows:
\[
L(z_1,z_0,\alpha,\lambda)=
C^{-1}(z_1,\lambda)\widetilde{L}(z_1,z_0,\alpha,\lambda)C(z_0,\lambda)=
\left(\begin{array}{cc} z_0-z_1-\alpha & (z_0-z_1)^2-\alpha^2 \\ \\ 
\lambda\alpha & z_0-z_1+\alpha \end{array}\right),
\]
which coincides with (\ref{C Lax}) up to a scalar factor 
(inessential when considering action of $L$ as a M\"obius transformation).

Since the shifted ratio equation possesses the three--leg form, there holds
the analog of Proposition \ref{quadgraph to Toda}. The corresponding systems 
of the Toda type on $\cG$, $\cG^*$ will be called the {\itbf multiplicative 
rational Toda systems}. Their equations (in notations of
Figs.\,\ref{flower},\ref{star}) read: 
\begin{equation}\label{Toda C}
\prod_{k=1}^n\frac{z_0-z_{2k}-\alpha_k+\alpha_{k+1}}
{z_0-z_{2k}+\alpha_k-\alpha_{k+1}}=1.
\end{equation}
As in the previous section, we can find the transition matrix across the 
edge $\ee_k=(v_0,v_{2k})\in E(\cG)$, i.e. along the edge
$\ee_k^*=(v_{2k-1},v_{2k+1})\in E(\cG^*)$. This time we start with the
``untilded'' matrices (\ref{C Lax}):
\begin{equation}\label{L cross-rat to Toda C}
\cL_k(\lambda)=L(z_{2k+1},z_0,\alpha_{k+1},\lambda)
L(z_0,z_{2k-1},\alpha_k,\lambda).
\end{equation}
\begin{proposition}\label{Lax cross-rat to Toda C}
Under the inverse gauge transformation (\ref{gauge Psi}), (\ref{A gauge}),
and up to the scalar determinant normalizing factor,
the matrix (\ref{L cross-rat to Toda C}) turns into
\begin{eqnarray}\label{Toda C Lax}
\lefteqn{\widetilde{\cL}(z_{2k},z_0,\lambda)=
A(z_{2k+1})\cL_k(\lambda)A^{-1}(z_{2k-1})=
\left(\begin{array}{cc} 1 & (\alpha_{k+1}-\alpha_k)\,\displaystyle\frac
{z_0+z_{2k}-\alpha_k-\alpha_{k+1}}{z_0-z_{2k}+\alpha_k-\alpha_{k+1}} \\ \\
0 & \displaystyle\frac{z_0-z_{2k}-\alpha_k+\alpha_{k+1}}
{z_0-z_{2k}+\alpha_k-\alpha_{k+1}}  \end{array}\right)+}\nonumber\\ 
                                                            \nonumber\\
& & +\displaystyle\frac{\lambda}{z_0-z_{2k}+\alpha_k-\alpha_{k+1}}
\left(\begin{array}{cc} \alpha_{k+1}z_{2k}-\alpha_kz_0 & 
(\alpha_k-\alpha_{k+1})(z_0z_{2k}-\alpha_k\alpha_{k+1}) \\ \\ 
\alpha_{k+1}-\alpha_k & \alpha_kz_{2k}-\alpha_{k+1}z_0\end{array}\right).
\end{eqnarray}
\end{proposition}
{\bf Proof.} The calculation of the $\lambda^0$--component of the matrix
(\ref{Toda C Lax}) is immediate, one has only to use the
three--leg equation (\ref{shifted rat 3-leg}), i.e.
\begin{equation}\label{shifted rat 3-leg gen}
\frac{z_0-z_{2k+1}+\alpha_{k+1}}{z_0-z_{2k+1}-\alpha_{k+1}}\cdot
\frac{z_0-z_{2k-1}-\alpha_k}{z_0-z_{2k-1}+\alpha_k}=
\frac{z_0-z_{2k}-\alpha_k+\alpha_{k+1}}{z_0-z_{2k}+\alpha_k-\alpha_{k+1}}.
\end{equation} 
The calculation of
the $\lambda$--component is also straightforward but somewhat more involved.
The main ingredient of this calculation is obtaining the following expression
for the entry 21 (which is used actually in all four entries):
\begin{equation}\label{Lax C aux}
\frac{\alpha_k}{z_{2k-1}-z_0-\alpha_k}\cdot
\frac{z_0-z_{2k+1}+\alpha_{k+1}}{z_0-z_{2k+1}-\alpha_{k+1}}+
\frac{\alpha_{k+1}}{z_0-z_{2k+1}-\alpha_{k+1}}=
\frac{\alpha_{k+1}-\alpha_k}{z_0-z_{2k}+\alpha_k-\alpha_{k+1}}.
\end{equation}
This is done as follows: one starts with the shifted ratio equation
\[
\alpha_k\,\frac{z_{2k}-z_{2k-1}-\alpha_{k+1}}{z_{2k-1}-z_0+\alpha_k}-
\alpha_{k+1}\,\frac{z_{2k+1}-z_{2k}+\alpha_k}{z_0-z_{2k+1}-\alpha_{k+1}}=0.
\]
This is rewritten as
\[
\alpha_k\,\frac{z_0-z_{2k}-\alpha_k+\alpha_{k+1}}{z_{2k-1}-z_0+\alpha_k}+
\alpha_{k+1}\,\frac{z_0-z_{2k}+\alpha_k-\alpha_{k+1}}
{z_0-z_{2k+1}-\alpha_{k+1}}=\alpha_{k+1}-\alpha_k.
\]
But the latter equation is equivalent to (\ref{Lax C aux}) due to the 
three--leg equation (\ref{shifted rat 3-leg gen}). \qed
\vspace{2mm}

The transition matrix for the type (C) Toda system found in \cite{A3} 
coincides with (\ref{Toda C Lax}) up to a similarity transformation 
with a constant matrix $\left(\begin{array}{cc} 1 & \lambda^{-1} \\ 0 & 1 
\end{array}\right)$.
\vspace{2mm}

Finally, we briefly discuss the dual to the shifted ratio system.
It appears, if one notices that the equation (\ref{shifted rat aux}) 
yields the existence of the function $Z:V(\cD)\mapsto{\Bbb C}$ 
such that on each face of $\cD$ there holds:
\[
Z_1-Z_0=\frac{1}{2}\,\log\frac{z_0-z_1+\alpha_1}{z_0-z_1-\alpha_1},\quad 
Z_2-Z_1=\frac{1}{2}\,\log\frac{z_1-z_2+\alpha_2}{z_1-z_2-\alpha_2},
\]
\[
Z_3-Z_2=\frac{1}{2}\,\log\frac{z_2-z_3+\alpha_1}{z_2-z_3-\alpha_1},\quad 
Z_0-Z_3=\frac{1}{2}\,\log\frac{z_3-z_0+\alpha_2}{z_3-z_0-\alpha_2}.
\] 
\begin{proposition}\label{C dual}
For any solution $z:V(\cD)\mapsto{\Bbb C}$ of the shifted ratio system on
$\cD$, the formula 
\begin{equation}\label{C dual edge}
Z(v_2)-Z(v_1)=\frac{1}{2}\,\log\frac{z(v_2)-z(v_1)+\alpha(\ee)}
{z(v_2)-z(v_1)-\alpha(\ee)}
\end{equation} 
(for any $\ee=(v_1,v_2)\in E(\cD)$) correctly defines a function 
$Z:V(\cD)\mapsto{\Bbb C}$ which is a solution
of the {\itbf hyperbolic cross--ratio system}
\begin{equation}\label{hyp cross-rat eq}
\frac{\sinh(Z_0-Z_1)\sinh(Z_2-Z_3)}
{\sinh(Z_1-Z_2)\sinh(Z_3-Z_0)}=\frac{\alpha_1}{\alpha_2}\;.
\end{equation}
\end{proposition}
\begin{remark}
{\rm The hyperbolic cross--ratio system is equivalent to the usual cross--ratio
system 
\[
\frac{(X_0-X_1)(X_2-X_3)}{(X_1-X_2)(X_3-X_0)}=\frac{\alpha_1}{\alpha_2}
\]
upon the change of variables $X_k=\exp(2Z_k)$. Restrictions of solutions
of the hyperbolic cross--ratio system to $V(\cG)$, $V(\cG^*)$ satisfy the
{\itbf additive hyperbolic Toda systems}, whose equations are
\begin{equation}\label{Toda B}
\sum_{k=1}^n{\rm (\alpha_k-\alpha_{k+1})coth}(z_0-z_{2k})=0.
\end{equation}
This system is also equivalent to the additive rational Toda system
\[
\sum_{k=1}^n\frac{\alpha_k-\alpha_{k+1}}{x_0-x_{2k}}=0
\]
under the change of variables $x_k=\exp(2z_k)$.
}
\end{remark}

\setcounter{equation}{0}
\section{Hyperbolic shifted ratio system and multiplicative hyperbolic 
Toda system}
\label{Sect cross-ratio D}

Finally, we turn to the analog of the cross--ratio system which is the most
general one.

\begin{definition}
Given a labelling of $E(\cD)$, the {\itbf hyperbolic shifted ratio system} on 
$\cD$ is the collection of equations for the function $z:V(\cD)\mapsto{\Bbb C}$, 
one equation per face of $\cD$, which read, in notations as on Fig. \ref{diamond
again} and $z_k=z(v_k)$:
\begin{equation}\label{hyp shifted rat eq 1}
\frac{\sinh(z_0-z_1+\alpha_1)\sinh(z_2-z_3+\alpha_1)}
{\sinh(z_1-z_2-\alpha_2)\sinh(z_3-z_0-\alpha_2)}=\frac{\sinh(2\alpha_1)}
{\sinh(2\alpha_2)}\;,
\end{equation}
or, equivalently,
\begin{equation}\label{hyp shifted rat eq 2}
\frac{\sinh(z_0-z_1-\alpha_1)\sinh(z_2-z_3-\alpha_1)}
{\sinh(z_1-z_2+\alpha_2)\sinh(z_3-z_0+\alpha_2)}=\frac{\sinh(2\alpha_1)}
{\sinh(2\alpha_2)}\;.
\end{equation}
\end{definition}

The hyperbolic shifted ratio system may be considered as a deformation of
the rational one, the corresponding limit from the former to the latter is
achieved by setting replace $z_k\mapsto\epsilon z_k$ and $\alpha_k\to 
\epsilon\alpha_k$ and then sending $\epsilon\to 0$.
\vspace{2mm}

We shall systematically use in this section the notations $x_k=\exp(2z_k)$, 
$c_k=\exp(2\alpha_k)$. In these notations the equations 
(\ref{hyp shifted rat eq 1}), (\ref{hyp shifted rat eq 2}) take the following
rational form:
\begin{equation}\label{hyp shifted rat eq 1 rat}
\frac{(c_1x_0-x_1)(c_1x_2-x_3)}{(x_1-c_2x_2)(x_3-c_2x_0)}=\frac{1-c_1^2}
{1-c_2^2}\;,
\end{equation}
and
\begin{equation}\label{hyp shifted rat eq 2 rat}
\frac{(x_0-c_1x_1)(x_2-c_1x_3)}{(c_2x_1-x_2)(c_2x_3-x_0)}=\frac{1-c_1^2}
{1-c_2^2}\;.
\end{equation}
\begin{proposition}\label{hyperbolic shifted ratio Lax}
The hyperbolic shifted ratio system on $\cD$ admits a zero curvature 
representation in the sense of Sect.\,\ref{Sect integrable systems on graphs}, 
with the transition matrices for $\ee=(v_0,v_1)$: 
\begin{equation}\label{D Lax}
L(\ee,\lambda)=L(x_1,x_0,c,\lambda)=
(1-\lambda(1-c^2))^{-1/2}\left(\begin{array}{cc}
1 & x_0-cx_1 \\ \\ \displaystyle\frac{\lambda(1-c^2)}{cx_0-x_1} & 
\displaystyle\frac{x_0-cx_1}{cx_0-x_1}  \end{array}
\right).
\end{equation}
Here $c=c(\ee)$, and $x_k=x(v_k)$.
\end{proposition}
{\bf Proof} is absolutely parallel to the proof of Proposition 
\ref{shifted ratio Lax}. \qed
\vspace{2mm}

Again, this statement may be derived from the {\itbf three--dimensional
consistency} of the shifted hyperbolic ratio system.
\begin{proposition}\label{hyp shifted rat 3-dim compatibility} 
The hyperbolic shifted ratio system is consistent on the three--dimensional 
quad--graph ${\bf D}$.
\end{proposition}
{\bf Proof} -- a direct computation. \qed
\vspace{2mm}

The derivation of a zero curvature representation from this fact is not
different from the previous two cases: we assign the label $\mu$ to all 
``vertical'' edges $(v,v')$ of $\bf D$, set $\nu=\exp(2\mu)$, and consider 
the hyperbolic shifted ratio equation on the ``vertical'' face 
$(v_0,v_1,v_1',v_0')$:
\[
\frac{(cx_0-x_1)(cx_1'-x_0')}{(x_1-\nu x_1')(x_0'-\nu x_0)}=\frac{1-c^2}
{1-\nu^2}\;.
\]
Solving for $x_1'$ in terms of $x_0'$, we find:
\[
x_1'=\widetilde{L}(x_1,x_0,c,\nu)[x_0'],
\]
where
\[
\widetilde{L}(x_1,x_0,c,\nu)=\left(\begin{array}{cc}
cx_0-x_1+\displaystyle\frac{1-c^2}{1-\nu^2}x_1 & 
-\displaystyle\frac{\nu(1-c^2)}{1-\nu^2}x_0x_1 \\ \\
\displaystyle\frac{\nu(1-c^2)}{1-\nu^2} & 
x_0-cx_1-\displaystyle\frac{1-c^2}{1-\nu^2}x_0\end{array}\right).
\]
Perform a gauge transformation
\begin{equation}\label{gauge Psi D}
\Psi(v)\mapsto D(x(v),\nu)\Psi(v),
\end{equation}
where
\begin{equation}\label{D gauge}
D(x,\nu)=\left(\begin{array}{cc} \nu^{-1} & \nu^{-1}x \\ \\ 0 & 1 
\end{array}\right)=\left(\begin{array}{cc} \nu^{-1} & 0 \\ \\ 0 & 1 
\end{array}\right)A(x).
\end{equation}
It leads to the following form of the transition matrices:
\[
L(x_1,x_0,c,\lambda)=
D^{-1}(x_1,\nu)\widetilde{L}(x_1,x_0,c,\nu)D(x_0,\nu)=
\left(\begin{array}{cc} cx_0-x_1 & (cx_0-x_1)(x_0-cx_1) \\ \\ 
\lambda(1-c^2) & x_0-cx_1\end{array}\right),
\]
where $\lambda=(1-\nu^2)^{-1}$. This coincides with (\ref{D Lax}) up to an 
inessential scalar factor.
\vspace{2mm}
 
Next, we turn to the derivation of the Toda type system related to the
hyperbolic shifted ratio system. As usual, the {\itbf three--leg form}
of the latter system is fundamentally important in this respect. It is
not difficult to check that (\ref{hyp shifted rat eq 1}) is equivalent to
the following equation:
\begin{equation}\label{hyp shifted rat 3-leg}
\frac{\sinh(z_0-z_1+\alpha_1)}{\sinh(z_0-z_1-\alpha_1)}\cdot
\frac{\sinh(z_0-z_3-\alpha_2)}{\sinh(z_0-z_3+\alpha_2)}=
\frac{\sinh(z_0-z_2+\alpha_1-\alpha_2)}{\sinh(z_0-z_2-\alpha_1+\alpha_2)}.
\end{equation} 
The rational form of the three--leg equation reads:
\begin{equation}\label{hyp shifted rat 3-leg rat}
\frac{(c_1x_0-x_1)}{(x_0-c_1x_1)}\cdot
\frac{(x_0-c_2x_3)}{(c_2x_0-x_3)}=
\frac{(c_1x_0-c_2x_2)}{(c_2x_0-c_1x_2)}\;.
\end{equation}
The three--leg form of the equations immediately yields the analog
of Proposition \ref{quadgraph to Toda}. The corresponding Toda type systems
on $\cG$, resp. $\cG^*$, read (in notations of Figs.\,\ref{flower},\ref{star}):
\begin{equation}\label{Toda D}
\prod_{k=1}^n\frac{\sinh(z_0-z_{2k}+\alpha_k-\alpha_{k+1})}
{\sinh(z_0-z_{2k}-\alpha_k+\alpha_{k+1})}=1,
\end{equation}
or in the rational form
\begin{equation}\label{Toda D rat}
\prod_{k=1}^n\frac{c_kx_0-c_{k+1}x_{2k}}{c_{k+1}x_0-c_kx_{2k}}=1.
\end{equation}
They will be called {\itbf multiplicative hyperbolic Toda systems}.

Turning to the zero curvature formulation of the multiplicative hyperbolic Toda 
system  on $\cG$, we find, as in the previous section, the transition matrix 
across the edge $\ee_k=(v_0,v_{2k})\in E(\cG)$, i.e. along the edge
$\ee_k^*=(v_{2k-1},v_{2k+1})\in E(\cG^*)$:
\begin{equation}\label{L cross-rat to Toda D}
\cL_k(\lambda)=L(x_{2k+1},x_0,c_{k+1},\lambda)L(x_0,x_{2k-1},c_k,\lambda).
\end{equation}
\begin{proposition}\label{Lax cross-rat to Toda D}
Under the inverse gauge transformation (\ref{gauge Psi}), (\ref{A gauge}),
and up to the scalar determinant normalizing factor,
the matrix (\ref{L cross-rat to Toda D}) turns into
\begin{eqnarray}\label{L cross-rat to Toda D gauge}
\lefteqn{\widetilde{\cL}(x_{2k},x_0,\lambda)=
A(x_{2k+1})\cL_k(\lambda)A^{-1}(x_{2k-1})=}\nonumber\\  \nonumber\\
& & =\left(\begin{array}{cc} (1-\lambda)+\lambda c_kc_{k+1}\displaystyle\frac
{c_kx_0-c_{k+1}x_{2k}}{c_{k+1}x_0-c_kx_{2k}} & 
(1-\lambda)\displaystyle\frac
{(c_{k+1}^2-c_k^2)x_0x_{2k}}{c_{k+1}x_0-c_kx_{2k}} \\ \\
\lambda\,\displaystyle\frac{c_{k+1}^2-c_k^2}{c_{k+1}x_0-c_kx_{2k}} & 
(1-\lambda)\displaystyle\frac{c_kx_0-c_{k+1}x_{2k}}
{c_{k+1}x_0-c_kx_{2k}}+\lambda c_kc_{k+1}  \end{array}\right). 
\end{eqnarray}
\end{proposition}
{\bf Proof} -- by a direct calculation. The key formula in this calculation is:
\begin{equation}\label{Lax D aux}
\frac{1-c_k^2}{x_0-c_kx_{2k-1}}\cdot
\frac{x_0-c_{k+1}x_{2k+1}}{c_{k+1}x_0-x_{2k+1}}-
\frac{1-c_{k+1}^2}{c_{k+1}x_0-x_{2k+1}}=
\frac{c_{k+1}^2-c_k^2}{c_{k+1}x_0-c_kx_{2k}}
\end{equation}
which is proved similarly to (\ref{Lax C aux}), based on the three--leg 
formula (\ref{hyp shifted rat 3-leg rat}). \qed
\vspace{2mm}

Again, the transition matrices for the multiplicative hyperbolic Toda systems
essentially equivalent to (\ref{L cross-rat to Toda D gauge}) were found in
\cite{A3}.
\vspace{2mm}

The multiplicative hyperbolic Toda lattice can be given a geometrical 
interpretation. For this aim consider, as at the end of Sect.
\ref{subsect patterns A}, a circle pattern with the combinatorics of a
cellular decomposition $\cG$, and extend it by the Euclidean centers of
the circles. So, the Euclidean radii of the circles $r:V(\cG^*)\mapsto
{\Bbb R}_+$ are defined. For any two neighboring circles $C_0$, $C_1$ of 
the pattern we use the notations as on Fig.\,\ref{two circles}, including 
the notation $\psi$ for the angle $z_0z_3z_2$. It is not difficult to express 
the angle $\psi$ in terms of the radii $r_0$, $r_1$ of the circles $C_0$, 
$C_1$ and their intersection angle $\phi$:
\begin{equation}
\exp(i\psi)=\frac{r_0+r_1\exp(-i\phi)}{r_0+r_1\exp(i\phi)}.
\end{equation}
For any circle $C_0$ of the pattern, denote by $C_k$, $k=1,2,\ldots,m$ 
its $m$ neighboring circles. Let $r_k$ be the radius of the circle $C_k$,
denote by $\psi_k$, $k=1,2,\ldots,m$ the angles $\psi$ corresponding to all 
pairs $(C_0,C_k)$, and denote by $\phi_k$, $k=1,2,\ldots,m$ the
intersection angles of $C_0$ with $C_k$. Obviously, for geometrical reasons 
we have:
\[
\prod_{k=1}^m \exp(i\psi_k)=1,
\]
therefore
\begin{equation}\label{identity for radii}
\prod_{k=1}^m \frac{r_0+r_k\exp(-i\phi_k)}{r_0+r_k\exp(i\phi_k)}=1.
\end{equation}
Now from Lemmas \ref{lemma on cross-ratios}, \ref{lemma on angles} there 
follows:
\begin{proposition}
Let $\cG$ be a cellular decomposition of a topological disc (open or closed),
and consider a circle pattern with the combinatorics of $\cG$. Suppose 
that the intersection angles of the circles satisfy the condition from
Proposition \ref{labelling for pattern}, namely:
\begin{itemize}
\item For each circle of the pattern the sum of its intersection 
angles $\phi_k$, $k=1,2,\ldots,m$ with all neighboring circles
of the pattern satisfies $2\sum_{k=1}^m\phi_k\equiv 0\pmod{2\pi}$.
\end{itemize}
Then there exists a labelling $c:E(\cD)\mapsto{\Bbb S}^1=
\{\theta\in{\Bbb C}:|\theta|=1\}$ depending only on the intersection angles,
such that the Euclidean radii of the circles $r:V(\cG^*)\mapsto{\Bbb R}_+$ 
satisfy the corresponding multiplicative hyperbolic Toda system on $\cG^*$:
\begin{equation}\label{Toda for radii}
\prod_{k=1}^m \frac{c_kr_0-c_{k+1}r_k}{c_{k+1}r_0-c_kr_k}=1.
\end{equation} 
\end{proposition}
{\bf Proof.} Intersection angles are naturally assigned to the edges from
$E(\cG)$. So, on Fig.\ref{two circles} $\phi=\phi(v_0,v_2)$. Comparing 
(\ref{identity for radii}) with (\ref{Toda for radii}),
we see that the problem in question is finding a labelling $c:E(\cD)\mapsto
{\Bbb S}^1$ such that, in the notations of Fig.\,\ref{diamond again},
\[
-\exp(i\phi(v_0,v_2))=\exp(i\pi-i\phi(v_0,v_2))=\frac{c_1}{c_2}.
\]
The end of the proof is exactly the same as that of Proposition 
\ref{labelling for pattern}. \qed
\vspace{2mm}

Finally, we discuss the duality for the hyperbolic shifted ratio system. 
As a basis for the duality there serves the formula
\begin{equation}\label{hyp shifted rat aux}
\frac{\sinh(z_2-z_1+\alpha_2)}{\sinh(z_2-z_1-\alpha_2)}\cdot
\frac{\sinh(z_1-z_0+\alpha_1)}{\sinh(z_1-z_0-\alpha_1)}=
\frac{\sinh(z_2-z_3+\alpha_1)}{\sinh(z_2-z_3-\alpha_1)}\cdot
\frac{\sinh(z_3-z_0+\alpha_2)}{\sinh(z_3-z_0-\alpha_2)},
\end{equation}
which is a simple consequence of (\ref{hyp shifted rat eq 1}), 
(\ref{hyp shifted rat eq 2}). As usual, the latter equation allows us to 
introduce the dual system:
\begin{proposition}\label{D dual}
For any solution $z:V(\cD)\mapsto{\Bbb C}$ of the hyperbolic shifted ratio 
system on $\cD$, the formula 
\begin{equation}\label{D dual edge}
Z(v_2)-Z(v_1)=\frac{1}{2}\,\log\frac{\sinh\Big(z(v_1)-z(v_2)+\alpha(\ee)\Big)}
{\sinh\Big(z(v_1)-z(v_2)-\alpha(\ee)\Big)}
\end{equation} 
(for any $\ee=(v_1,v_2)\in E(\cD)$) correctly defines a function 
$Z:V(\cD)\mapsto{\Bbb C}$ which is another solution of the same
hyperbolic shifted ratio system.
\end{proposition}
In the notation $X(v)=\exp(2Z(v))$, the latter formula reads:
\begin{equation}\label{D dual edge rat}
\frac{X(v_2)}{X(v_1)}=\frac{c(\ee)x(v_1)-x(v_2)}{x(v_1)-c(\ee)x(v_2)}.
\end{equation} 
In particular, on each face of $\cD$ we have:
\[
\frac{X_1}{X_0}=\frac{c_1x_0-x_1}{x_0-c_1x_1},\quad 
\frac{X_2}{X_1}=\frac{c_2x_1-x_2}{x_1-c_2x_2},\quad
\frac{X_3}{X_2}=\frac{c_1x_2-x_3}{x_2-c_1x_3},\quad 
\frac{X_0}{X_3}=\frac{c_2x_3-x_0}{x_3-c_2x_0}.
\] 
The first and the fourth of these equations together with formula
(\ref{hyp shifted rat 3-leg rat}) yield:
\begin{equation}\label{Toda D dual}
\frac{X_1}{X_3}=\frac{c_1x_0-c_2x_2}{c_2x_0-c_1x_2}
\end{equation}
This shows that, like for the additive rational Toda systems, a solution of 
the multiplicative hyperbolic Toda system on $\cG$ determines the dual 
solution of the multiplicative hyperbolic Toda system on $\cG^*$ almost 
uniquely (up to a constant factor, in terms of the variables $x,X$, or up 
to an additive constant, in terms of the variables $z,Z$).
\vspace{2mm}

One can also consider the function $w: V(\cD)\mapsto \widehat{\Bbb C}$
which coincides with $x$ on $V(\cG)$ and with $X$ on $V(\cG^*)$. Then 
formula (\ref{Toda D dual}) defines a system of equations on 
the diamond, reading on the face $(v_0,v_1,v_2,v_3)$ as
\begin{equation}\label{Hirota}
\frac{w_1}{w_3}=\frac{c_1w_0-c_2w_2}{c_2w_0-c_1w_2}.
\end{equation}
This is a well--defined equation on the quad--graph, which in the simplest
possible situation (square lattice, constant labelling) was introduced in
\cite{H2} and therefore will be called the {\it Hirota equation on} $\cD$.
It appeared also in \cite{NQC} under the name of ``discrete MKdV''.
Actually, when considered on a single face, this equation represents nothing 
but the famous result of Bianchi on the permutability of B\"acklund
transformations for the sine-Gordon equation. See a discussion of the role
of this equation (on a square lattice) for the geometrical and analytic
discretization of surfaces of a constant negative Gaussian curvature in
\cite{BP2}. 
 
The equation (\ref{Hirota}) is in the three--leg form, and restrictions
of its solutions to $V(\cG)$ and to $V(\cG^*)$ are solutions of the
multiplicative hyperbolic Toda systems on the corresponding graphs.
Also this equation possesses the property of the three--dimensional
consistency, which yields integrability: assign the label 
$\mu$ to all ``vertical'' edges $(v,v')$ of $\bf D$, and consider the 
Hirota equation on the ``vertical'' face $(v_0,v_1,v_1',v_0')$:
\[
\frac{w_0'}{w_1}=\frac{cw_1'-\mu w_0}{\mu w_1'-cw_0}.
\]
Solving for $w_1'$ in terms of $w_0'$, we find:
\[
w_1'=\frac{cw_0w_0'-\mu w_0w_1}{\mu w_0'-cw_1}=
\left(\begin{array}{cc}
cw_0 & -\mu w_0w_1 \\ 
\mu & -cw_1\end{array}\right)[w_0']=L(w_1,w_0,c,\mu)[w_0'],
\]
where (after normalizing the determinant)
\begin{equation}\label{Hirota L}
L(w_1,w_0,c,\mu)=(\mu^2-c^2)^{-1/2}\left(\begin{array}{cc}
c\left(\displaystyle\frac{w_0}{w_1}\right)^{1/2} & -\mu(w_0w_1)^{1/2} \\ 
\mu(w_0w_1)^{-1/2} & -c\left(\displaystyle\frac{w_1}{w_0}\right)^{1/2}
\end{array}\right).
\end{equation}
This is the well--known transition matrix for the Hirota equation \cite{H2},
\cite{FV}.

We remark also that there holds the diagram like that of Fig.\,\ref{diagram}:
with the replacement of the cross-ratio system by the hyperbolic shifted ratio
system, of the additive rational Toda systems by the multiplicative hyperbolic
Toda systems, and of the discrete KdV equation by the Hirota equation.

\setcounter{equation}{0}
\section{Conclusions}
\label{Sect conclusions}

The present paper is devoted to the study of discrete integrable systems
on arbitrary graphs. We considered several fundamental examples supporting
our viewpoint that integrable systems on quad-graphs play here a 
prominent role. In particular, the recently introduced systems of the Toda
type or arbitrary graphs find their natural and simple explanation on this
way. Let us indicate several points that were left open in this paper, and
some directions for the further progress.

1) We restricted ourselves here only to {\it local problems} of the theory
of integrable systems. Therefore, we considered only systems on graphs having
the combinatorics of a cellular decomposition of a topological disc. We intend
to address global problems for integrable systems on graphs embedded in 
arbitrary Riemann surfaces in our future work. This has important 
applications to the discrete differential geometry and the discrete
complex analysis.

2) Although we have unveiled several mysteries of integrable systems here
(in particular, let us mention the derivation of zero curvature representation
from the three--dimensional consistency of discrete systems), there remain
several phenomena which still wait for their explanation. For example, 
a deeper reasons for the existense of the {\it three--leg form} for all our 
equations are still unclear. Also the role of the duality and its 
interrelations with integrability calls for a better understanding.

3) In all our examples the basic equation on a face of the quad--graph
may be written as (cf. Fig.\,\ref{diamond again})
\[
\Phi(z_0,z_1,z_2,z_3,\alpha_1,\alpha_2)=0,
\]
where the function $\Phi$ is a polynomial (of degree $\le 2$), linear in each
of the arguments $z_k$. As a result, this equation may be solved uniquely for
each of $z_k$, the result being a linear--fractional (M\"obius) transformation
with respect to any of the remaining variables. As we have seen, the
three--dimensional permutability of M\"obius transformations of this kind
lies in the heart of the integrability. A further important class of 
integrable four--point equations, with functions $\Phi$ being polynomials of
degree $\le 4$, linear in each of the arguments, was found in \cite{A1} and
identified as a superposition formula for B\"acklund transformations of
the Krichever--Novikov equation. Thus, these integrable systems
are related to an elliptic curve, where the spectral parameter of the
Krichever--Novikov equation naturally lives \cite{KN1}.  In a subsequent 
publication \cite{AS} it will be shown that the corresponding equations on 
quad-graphs still possess the three--leg form (after parametrizing variables 
by elliptic $\sigma$--functions, $z_k=\sigma(u_k)$):
\begin{eqnarray*}
\lefteqn{\frac{\sigma(u_1-u_0+\alpha_1)\sigma(u_1+u_0-\alpha_1)}
{\sigma(u_1-u_0-\alpha_1)\sigma(u_1+u_0+\alpha_1)}\cdot
\frac{\sigma(u_3-u_0-\alpha_2)\sigma(u_3+u_0+\alpha_2)}
{\sigma(u_3-u_0+\alpha_2)\sigma(u_3+u_0-\alpha_2)}=}\\
& & =\frac{\sigma(u_2-u_0+\alpha_1-\alpha_2)\sigma(u_2+u_0-\alpha_1+\alpha_2)}
{\sigma(u_2-u_0-\alpha_1+\alpha_2)\sigma(u_2+u_0+\alpha_1-\alpha_2)}.
\end{eqnarray*}
As one of the consequences, we will find the {\it elliptic Toda system} on 
an arbitrary cellular decomposition $\cG$. In the notations of Fig.\,\ref{star} 
it reads:
\[
\prod_{k=1}^n
\frac{\sigma(u_{2k}-u_0+\alpha_k-\alpha_{k+1})
\sigma(u_{2k}+u_0-\alpha_k+\alpha_{k+1})}
{\sigma(u_{2k}-u_0-\alpha_k+\alpha_{k+1})
\sigma(u_{2k}+u_0+\alpha_k-\alpha_{k+1})}=1.
\]
We remark that in the degenerate case when $\sigma$--functions are replaced
by $\sinh$, the latter equation is again related to circle patterns with the
combinatorics of $\cG$: the radii of the circles calculated in the
non--Euclidean (hyperbolic) geometry always deliver a solution to such a 
system on $\cG^*$.

As the simplest particular case of this system (regular square lattice, 
constant labelling in each of the two directions), we will find \cite{AS}:
\begin{eqnarray*}
\lefteqn{\frac{\sigma(\wx_j-x_j+ha)}{\sigma(\wx_j-x_j-ha)}\cdot
\frac{\sigma(x_j+\wx_j-ha)}{\sigma(x_j+\wx_j+ha)}\cdot
\frac{\sigma(x_j-\undertilde{x}_j-ha)}{\sigma(x_j-\undertilde{x}_j+ha)}\cdot
\frac{\sigma(x_j+\undertilde{x}_j-ha)}{\sigma(x_j+\undertilde{x}_j+ha)}\cdot}
 \\
&&\frac{\sigma(x_j-x_{j-1}+ha)}{\sigma(x_j-x_{j-1}-ha)}\cdot
\frac{\sigma(x_j+x_{j-1}+ha)}{\sigma(x_j+x_{j-1}-ha)}\cdot
\frac{\sigma(x_{j+1}-x_j-ha)}{\sigma(x_{j+1}-x_j+ha)}\cdot
\frac{\sigma(x_{j+1}+x_j+ha)}{\sigma(x_{j+1}+x_j-ha)}
=1. 
\end{eqnarray*}
Here the index $j$ corresponds to one lattice direction of the regular square
lattice, and for the second lattice direction we use an index--free notation, 
so that the variables $x_j$
correspond to one and the same value of the second lattice coordinate $t\in
h{\Bbb Z}$, the tilded variables $\wx_j$ correspond to the value $t+h$, and
subtilded variables $\undertilde{x}_j$ correspond to the value $t-h$. In this
interpretation, the above equation serves as a time discretization of the
so called {\it elliptic Toda lattice} \cite{K1},
\[
\ddot{x}_j=(\dot{x}_j^2-a^2)\Big(\zeta(x_j-x_{j-1})+\zeta(x_j+x_{j-1})
-\zeta(x_{j+1}-x_j)+\zeta(x_{j+1}+x_j)-2\zeta(2x_j)\Big).
\] 
Similarly, if the cellular decomposition $\cG$ is generated by the regular
triangular lattice, the resulting discrete system may be considered as a time
discretization of a novel integrable lattice which is a sort of 
a ``relativistic'' generalization of the elliptic Toda lattice, cf. \cite{A2}. 
So, we will see that the root of integrability of all
these discrete and continuous systems still lies in the
three--dimensional permutability of certain M\"obius transformations.

4) In the light of above, there appears a meaningful problem of finding and 
classifying {\it all}\, equations $\Phi(z_0,z_1,z_2,z_3,\alpha_1,\alpha_2)=0$ 
with analogous properties, 
i.e. uniquely solvable with respect to each argument, and leading 
to transformations with the three--dimensional compatibility.
We plan to address this problem in our future research.  

5) Finally, a very intriguing problem is a generalization of the results
described above to the case of spectral parameter belonging to higher genus
algebraic curves, e.g. finding the discrete counterparts of the Hitchin
systems in Krichever's formulation \cite{K2}.

\section{Acknowledgements}
The research of the authors is partially supported by the Deutsche
Forschungsgemeinschaft in the frame of SFB288 ``Differential Geometry
and Quantum Physics''. We would like to thank Tim Hoffmann for the
collaboration and useful discussions, where some of the ideas of this paper
were born, in particular the idea of factorizing the transition matrices
for Toda type systems into the product of transition matrices for cross-ratio
type systems. We are grateful also to Vsevolod Adler for a useful
correspondence and for showing his results prior to publication.
 

\end{document}